\RequirePackage[hyphens]{url}
\documentclass[10pt,twocolumn,reprint]{revtex4-1}
\usepackage{lmodern}
\usepackage[T1]{fontenc}
\usepackage[latin9]{inputenc}
\usepackage{color}
\usepackage{array}
\usepackage{booktabs}
\usepackage{amsmath}
\usepackage{amssymb}
\usepackage{graphicx}

\usepackage[unicode=true,pdfusetitle,
 bookmarks=true,bookmarksnumbered=false,bookmarksopen=false,
 breaklinks=true,pdfborder={0 0 0},pdfborderstyle={},backref=false,colorlinks=true]
 {hyperref}

\makeatletter

\providecommand{\tabularnewline}{\\}

\usepackage{algcompatible}
\usepackage{placeins}
\usepackage{amsmath}
\renewcommand\toprule{\hline\hline}
\renewcommand\bottomrule{\hline\hline}

\makeatother

\begin{document}

\title{Optimizing short stabilizer codes for asymmetric channels}

\author{Alex Rigby}
\email{alex.rigby@utas.edu.au}

\author{JC Olivier}

\author{Peter Jarvis}

\affiliation{College of Sciences and Engineering, University of Tasmania, Hobart,
Tasmania 7005, Australia }
\begin{abstract}
For a number of quantum channels of interest, phase-flip errors occur
far more frequently than bit-flip errors. When transmitting across
these asymmetric channels, the decoding error rate can be reduced
by tailoring the code used to the channel. However, analyzing the
performance of stabilizer codes on these channels is made difficult
by the \#P-completeness of optimal decoding. To address this, at least
for short codes, we demonstrate that the decoding error rate can be
approximated by considering only a fraction of the possible errors
caused by the channel. Using this approximate error rate calculation,
we extend a recent result to show that there are a number of $[[5\leq n\leq12,1\leq k\leq3]]$
cyclic stabilizer codes that perform well on two different asymmetric
channels. We also demonstrate that an indication of a stabilizer code's
error rate is given by considering the error rate of a classical binary
code related to the stabilizer. This classical error rate is far less
complex to calculate, and we use it as the basis for a hill climbing
algorithm, which we show to be effective at optimizing codes for asymmetric
channels. Furthermore, we demonstrate that simple modifications can
be made to our hill climbing algorithm to search for codes with desired
structure requirements. 
\end{abstract}
\maketitle

\section{Introduction }

Quantum codes can be employed to protect quantum information against
the effects of a noisy channel. Of particular note are the stabilizer
codes, which are defined by a stabilizer $\mathcal{S}$ that is an
Abelian subgroup of the $n$-qubit Pauli group $\mathcal{P}_{n}$
\citep{gottesman1997stabilizer}. An $[[n,k]]$ stabilizer code encodes
the state of a $k$-qubit system in that of an $n$-qubit system;
that is, it is a subspace $\mathcal{Q}\subseteq(\mathbb{C}^{2})^{\otimes n}$
of dimension $2^{k}$. For a Pauli channel, an error $E$ acting on
the code is also an element of $\mathcal{P}_{n}$, with the component
acting on any given qubit being $I$ with probability $p_{I}$, $X$
with probability $p_{X}$, $Y$ with probability $p_{Y}$, or $Z$
with probability $p_{Z}$. Most stabilizer codes are implicitly designed
for good decoding performance (that is, a low decoding error rate)
on the depolarizing channel, where $p_{X}=p_{Y}=p_{Z}$. This is achieved
by ensuring that the code has large distance $d$, which is the weight
of the lowest weight error that yields a trivial syndrome while having
a nontrivial effect on the code. However, for a number of channels
of physical interest, $Z$ errors occur far more frequently than $X$
errors \citep{evans2007error,ioffe2007asymmetric}. For these channels,
better decoding performance can be achieved by using codes that are
tailored to the channel \citep{robertson2017tailored,tuckett2018ultrahigh}. 

In this paper, our focus is on the construction of stabilizer codes
for two different asymmetric channels. The first of these is the biased
$XZ$ channel, for which the $X$ and $Z$ components of an error
occur independently at different rates. The second is a Pauli approximation
of the combined amplitude damping (AD) and dephasing channel. Both
of these channels have two degrees of freedom, which means that the
values of $p_{X}$, $p_{Y}$, and $p_{Z}$ can be defined via the
total error probability $p=p_{X}+p_{Y}+p_{Z}$ and bias $\eta=p_{Z}/p_{X}$
\citep{robertson2017tailored,sarvepalli2009asymmetric}. A well-studied
approach to constructing codes for asymmetric channels is to restrict
consideration to Calderbank-Shor-Steane (CSS) codes \citep{calderbank1996good,steane1996multiple},
which can be designed to have separate $X$ and $Z$ distances $d_{X}$
and $d_{Z}$ (typically $d_{Z}>d_{X}$) \citep{sarvepalli2009asymmetric,aly2008asymmetric,ezerman2013css,wang2010asymmetric,la2013asymmetric,guenda2013symmetric}.
We wish to take a more direct approach to the problem by actually
determining the decoding error rates of the codes we construct (this
also allows us to meaningfully consider non-CSS codes). However, to
do this, we have to overcome the \#P-completeness of stabilizer decoding
\citep{iyer2015hardness}, which stems from the equivalence of errors
up to an element of the stabilizer. To achieve this, at least for
short codes (that is, codes with small $n$), we first demonstrate
that the error rate of an optimal decoder can be approximated by considering
only a small subset $\mathcal{E}$ of the $4^{n}$ possible Pauli
errors. We derive a bound on the relative error in this approximation,
and we demonstrate that the independence of error components can be
exploited to construct $\mathcal{E}$ without having to enumerate
all possible errors. We also show that the performance of a classical
$[2n,n+k]$ binary linear code associated with the stabilizer \citep{calderbank1997quantum,gottesman1997stabilizer}
gives an indication of the stabilizer code's performance (note that
whenever we mention a code's performance or error rate, we are referring
to that of the associated decoder). It is several orders of magnitude
faster to calculate this classical error rate, and we show that it
can itself be approximated using a limited error set.

We have a particular focus on cyclic codes, which are stabilizer codes
based on classical self-orthogonal additive cyclic $\mathrm{GF}(4)$
codes \citep{calderbank1996quantum,huffman2007additive,huffman2008additive}
{[}where $\mathrm{GF}(q)$ is the $q$-element finite field{]}. This
is motivated by the recent result of Ref. \citep{robertson2017tailored},
where it was shown that a $[[7,1]]$ cyclic code performs near-optimally
compared to $10\,000$ randomly constructed codes on the biased $XZ$
channel for a range of error probabilities and biases. We extend this
result by enumerating the $[[5\leq n\leq12,1\leq k\leq3]]$ cyclic
codes and making use of our approximate error rate calculation. In
particular, we demonstrate that there are a number of cyclic codes
that perform well compared to the best of $10\,000$ randomly constructed
codes for both the biased $XZ$ and AD channels across a range of
$p$ and $\eta$ values. In some cases, such as $[[n\geq9,1]]$ codes
for the biased $XZ$ channel, the best cyclic codes significantly
outperform the best of the random codes constructed. To improve on
the poor performance of the random search, we demonstrate the effectiveness
of a simple hill climbing algorithm that attempts to optimize the
performance of the classical binary code associated with a stabilizer.
We also show that by modifying the mutation operation employed by
this hill climbing algorithm, we can effectively search for codes
with desired structure. In particular, we show that we can search
for codes with weight-four generators, CSS codes, and linear codes. 

The paper is organized as follows. Sec. \ref{sec:Background} gives
an overview of classical codes, asymmetric quantum channels, and stabilizer
codes. In Sec. \ref{sec:Approximate-decoding}, we detail our methods
for calculating approximate error rates. In Sec. \ref{sec:Code-performance},
we demonstrate the performance of cyclic codes, outline our hill climbing
search algorithm, and show its effectiveness. The paper is concluded
in Sec. \ref{sec:Conclusion}.

\section{Background\label{sec:Background}}

\subsection{Classical codes\label{subsec:Classical-codes}}

A classical channel $\Phi$ maps a set of inputs $\mathcal{A}_{x}$
to a set of outputs $\mathcal{A}_{y}$. We are interested in the case
where $\mathcal{A}_{x}=\mathcal{A}_{y}=\mathrm{GF}(q)$, for which
the action of the channel is given by
\begin{equation}
\Phi(x)=x+e=y,
\end{equation}
where $x\in\mathrm{GF}(q)$ is the channel input, $y\in\mathrm{GF}(q)$
is the channel output, and $e\in\mathrm{GF}(q)$ is an error (or noise)
symbol that occurs with probability $P(e)$. $\Phi$ is called symmetric
if $P(0)=1-p$ and $P(e_{i})=p/(q-1)$ for $e_{i}\neq0$. The noise
introduced by the channel can be protected against using a code $\mathcal{C}\subseteq\mathrm{GF}(q)^{n}$,
whose elements are called codewords. The effect of the combined channel
$\Phi^{n}$, which is comprised of $n$ copies of $\Phi$, on some
codeword $\boldsymbol{x}\in\mathcal{C}$ is
\begin{equation}
\Phi^{n}(\boldsymbol{x})=\boldsymbol{x}+\boldsymbol{e}=\boldsymbol{y},
\end{equation}
where $\boldsymbol{y}\in\mathrm{GF}(q)^{n}$ is the channel output
and $\boldsymbol{e}\in\mathrm{GF}(q)^{n}$ is an error ``vector.''
Assuming that error components occur independently, the probability
of $\boldsymbol{e}=(e_{1},\dots,e_{n})$ occurring is
\begin{equation}
P(\boldsymbol{e})=\prod_{i=1}^{n}P(e_{i}),\label{eq:classical error probability}
\end{equation}
where $P(e_{i})$ is the probability of the error symbol $e_{i}$
occurring on $\Phi$. It follows that for a symmetric channel, the
probability of an error $\boldsymbol{e}$ occurring depends only on
its weight $w(\boldsymbol{e})$, which is the number of nonzero components
from which it is comprised. The distance $d$ of a code is the weight
of the lowest weight error mapping one codeword to another. The (minimum)
weight $w(\mathcal{C})$ of a code $\mathcal{C}$ is simply the weight
of the lowest weight codeword it contains.

A code is called additive if it forms a group (under addition) and
linear if it forms a vector space. Such codes can be described by
a generator matrix

\begin{equation}
G^{T}=\left(\begin{array}{ccc}
\boldsymbol{b}_{1} & \cdots & \boldsymbol{b}_{k}\end{array}\right),
\end{equation}
where $\mathcal{B}=\{\boldsymbol{b}_{1},\dots,\boldsymbol{b}_{k}\}$
is either a generating set or basis, respectively (note that we consider
codewords as column vectors). A linear code can also be defined as
the kernel of a $\mathrm{GF}(q)$ parity-check matrix $H$; that is,
\begin{equation}
\mathcal{C}=\{\boldsymbol{x}\in\mathrm{GF}(q)^{n}:H\boldsymbol{x}=\boldsymbol{0}\}.
\end{equation}
If $H$ has $m$ rows, then $\dim(\mathcal{C})=k\geq n-m$, with equality
when $H$ is full rank. For a linear code, the errors mapping one
codeword to another are themselves codewords; therefore, it follows
that the distance of a linear code $\mathcal{C}$ is simply $d=w(\mathcal{C})$.
A linear code of length $n$ with dimension $k$ and distance $d$
is called an $[n,k]_{q}$ or $[n,k,d]_{q}$ code (the $q$ is typically
omitted for binary codes, where $q=2$). More generally, a length-$n$
code of size $|\mathcal{C}|=K$ and distance $d$ is called an $(n,K)_{q}$
or $(n,K,d)_{q}$ code.

The dual code of some $\mathcal{C}\subseteq\mathrm{GF}(q)^{n}$ with
respect to the inner product $\langle\cdot,\cdot\rangle:\mathrm{GF}(q)^{n}\times\mathrm{GF}(q)^{n}\rightarrow\mathrm{GF}(q)$
is
\begin{equation}
\mathcal{C}^{\perp}=\{\boldsymbol{c}\in\mathrm{GF}(q)^{n}:\langle\boldsymbol{c},\boldsymbol{x}\rangle=0\,\,\forall\,\,\boldsymbol{x}\in\mathcal{C}\}.
\end{equation}
$\mathcal{C}^{\perp}$ is the annihilator of $\mathcal{C}$ and is
therefore a linear code. If $\mathcal{C}^{\perp}\subseteq\mathcal{C}$,
then $\mathcal{C}$ is called dual containing; if $\mathcal{C}\subseteq\mathcal{C}^{\perp}$,
then $\mathcal{C}$ is called self-orthogonal; and if $\mathcal{C}^{\perp}=\mathcal{C}$,
then $\mathcal{C}$ is called self-dual. Note that if $\mathcal{C}$
is dual containing, then $\mathcal{C}^{\perp}$ is self-orthogonal
and vice versa. Unless otherwise specified, the dual code is with
respect to the Euclidean inner product 
\begin{equation}
\langle\boldsymbol{c},\boldsymbol{x}\rangle=\boldsymbol{c}\cdot\boldsymbol{x}=\sum_{i=1}^{n}c_{i}x_{i}.
\end{equation}
In this case, if $\mathcal{C}$ is linear with generator matrix $G$,
then a necessary and sufficient condition for $\boldsymbol{c}\in\mathcal{C}^{\perp}$
is \textbf{$G\boldsymbol{c}=\boldsymbol{0}$}; that is, a generator
matrix for $\mathcal{C}$ is a parity-check matrix for $\mathcal{C}^{\perp}$.
Conversely, if $H$ is a parity-check matrix for $\mathcal{C}$, then
it is a generator matrix for $\mathcal{C}^{\perp}$.

A decoder uses the output of a channel to infer its input. For a linear
code, this inference can be aided by the syndrome 
\begin{equation}
\boldsymbol{z}=H\boldsymbol{y}=H(\boldsymbol{x}+\boldsymbol{e})=H\boldsymbol{e}.\label{eq:classical syndrome definition}
\end{equation}
As channel outputs that differ only by a codeword yield the same syndrome,
the $q^{n-k}$ possible syndromes can be associated with the cosets
of $\mathrm{GF}(q)^{n}/\mathcal{C}$. Given some syndrome measurement
$\boldsymbol{z}$, an optimal maximum a posteriori (MAP) decoder will
then return the most probable error
\begin{equation}
\hat{\boldsymbol{e}}_{\boldsymbol{z}}=\underset{\boldsymbol{e}\in\mathrm{GF}(q)^{n}}{\mathrm{argmax}}P(\boldsymbol{e}|\boldsymbol{z})\label{eq:syndrome decoder}
\end{equation}
in the corresponding coset. The channel input can then be inferred
as $\hat{\boldsymbol{x}}=\boldsymbol{y}-\hat{\boldsymbol{e}}_{\boldsymbol{z}}$.
If $\hat{\boldsymbol{e}}=\hat{\boldsymbol{e}}_{\boldsymbol{z}}$ (and
hence $\hat{\boldsymbol{x}}=\boldsymbol{x}$), then decoding is successful;
otherwise, a decoding error has occurred. The probability of such
a decoding error, called the frame error rate (FER), is simply
\begin{equation}
F=1-\sum_{\boldsymbol{z}\in\mathrm{GF}(q)^{n-k}}P(\hat{\boldsymbol{e}}_{z}).\label{eq:classical FER}
\end{equation}
Unfortunately, even in the simple case of a binary code operating
on the binary symmetric channel (a symmetric channel with $q=2$),
this decoding problem can be shown to be NP-complete \citep{berlekamp1978inherent}.
This complicates the design of highly performant codes (that is, codes
yielding a low FER). In practice, when designing codes for symmetric
channels, the simpler goal of achieving a large distance is typically
settled for. This is motivated by the fact that for low-distance codes,
there are many errors in each coset $\hat{\boldsymbol{e}}_{z}+\mathcal{C}$
with weight, and hence probability, similar to $\hat{\boldsymbol{e}}_{z}$,
which leads to a high FER according to Eq. (\ref{eq:classical FER})
(see Sec. II A of Ref. \citep{PhysRevA.100.012330} for a more detailed
discussion). 

Two codes $\mathcal{C}$ and $\mathcal{C}'$ are called permutation
equivalent if they are the same up to a relabeling of coordinates.
Permutation-equivalent codes share a large number of properties including
length, size, and distance; furthermore, they yield the same FER for
channels where the error components are independently and identically
distributed. While there are more general notions of code equivalence,
whenever we say that two codes are equivalent, we mean that they are
permutation equivalent in this paper. Furthermore, if some family
(set) of codes $\{\mathcal{C}_{1},\dots,\mathcal{C}_{N}\}$ can be
split into $M$ equivalence classes (according to permutation equivalence),
then we simply say that $M$ of the codes are inequivalent.

\subsection{Cyclic codes\label{subsec:Cyclic-codes}}

Cyclic codes are those for which a cyclic shift of any codeword is
also a codeword; that is, for a cyclic code $\mathcal{C}$, if $(c_{0},c_{1},\dots,c_{n-1})\in\mathcal{C}$,
then it is also the case that $(c_{n-1},c_{0},\dots,c_{n-2})\in\mathcal{C}$
(note that to be consistent with standard convention, we index the
codewords of cyclic codes from zero in this section). If $\mathcal{C}$
is linear, then it has a convenient description through the mapping
\begin{equation}
\boldsymbol{c}=(c_{0},c_{1},\dots,c_{n-1})\leftrightarrow c_{0}+c_{1}x+\dots+c_{n-1}x^{n-1}=c(x)
\end{equation}
of codewords to polynomials in $\mathrm{GF}(q)[x]$. Cyclic shifts
of codewords correspond to a multiplication by $x$ taken modulo $x^{n}-1$;
that is, $(c_{n-1},c_{0},\dots,c_{n-2})\leftrightarrow xc(x)\,(\mathrm{mod}\,x^{n}-1)$.
As $\mathcal{C}$ is linear, $r(x)c(x)\,(\mathrm{mod}\,x^{n}-1)$
is a codeword for any $r(x)\in\mathrm{GF}(q)[x]$, from which it follows
that $\mathcal{C}$ corresponds to an ideal $I_{\mathcal{C}}\in\mathrm{GF}(q)[x]/(x^{n}-1)$.
Any such ideal is principal and is generated by a unique monic polynomial
of minimal degree $g(x)\in I_{\mathcal{C}}$ that is a factor of $x^{n}-1$
\citep{macwilliams1977theory}; through slight abuse of notation,
we write $\mathcal{C}=\langle g(x)\rangle$. $\mathcal{C}$ has dimension
$k=n-\deg(g)$ and has a generator matrix
\begin{equation}
G=\left(\begin{array}{ccccc}
g_{0} & \cdots & g_{n-k} &  & 0\\
 & \ddots & \ddots & \ddots\\
0 &  & g_{0} & \cdots & g_{n-k}
\end{array}\right).
\end{equation}
Furthermore, a parity-check matrix 
\begin{equation}
H=\left(\begin{array}{ccccc}
h_{k} & \cdots & h_{0} &  & 0\\
 & \ddots & \ddots & \ddots\\
0 &  & h_{k} & \cdots & h_{0}
\end{array}\right)
\end{equation}
is given in terms of the check polynomial $h(x)=(x^{n}-1)/g(x)$.
It follows that the dual code $\mathcal{C}^{\perp}$ is also cyclic
and is generated by $x^{k}h(x^{-1})$.

In the quantum setting, we are particularly interested in codes over
$\mathrm{GF}(4)=\{0,1,\omega,\omega^{2}=\bar{\omega}\}$ that are
self-orthogonal with respect to the trace inner product (this will
be explained further in Sec. \ref{subsec:Stabilizer-codes}). Note
that the trace inner product of $\boldsymbol{a},\boldsymbol{b}\in\mathrm{GF}(4)^{n}$
is
\begin{equation}
\boldsymbol{a}*\boldsymbol{b}=\mathrm{tr}(\boldsymbol{a}\cdot\bar{\boldsymbol{b}})=\mathrm{tr}\left(\sum_{i=1}^{n}a_{i}\bar{b_{i}}\right),\label{eq:trace inner prod}
\end{equation}
where $\bar{0}=0$, $\bar{1}=1$, $\bar{\omega}=\omega^{2}$, and
$\bar{\omega^{2}}=\omega$; and $\mathrm{tr}(x)=x+\bar{x}$ (that
is, $\mathrm{tr}(0)=\mathrm{tr}(1)=0$ and $\mathrm{tr}(\omega)=\mathrm{tr}(\bar{\omega})=1$).
A linear cyclic $\mathrm{GF}(4)$ code $\mathcal{C}=\langle g(x)\rangle$
is self-orthogonal if and only if $g(x)g^{\dagger}(x)\equiv0\,(\mathrm{mod}\,x^{n}-1)$
\citep{calderbank1996quantum}, where
\begin{equation}
g^{\dagger}(x)=\bar{g}_{0}+\sum_{j=1}^{n-1}\bar{g}_{n-j}x^{j}.
\end{equation}

More generally, an $(n,2^{k})_{4}$ additive cyclic code $\mathcal{C}$
has two generators \citep{calderbank1996quantum,huffman2007additive,huffman2008additive}.
Following the formulation of Ref. \citep{calderbank1996quantum},
$\mathcal{C}=\langle\omega p(x)+q(x),r(x)\rangle$ where $p(x),q(x),r(x)\in\mathrm{GF}(2)[x]$;
$p(x)$ and $r(x)$ are factors of $x^{n}-1$; and $r(x)$ is also
a factor of $q(x)(x^{n}-1)/p(x)$. In general, the choice of generators
is not unique; however, any other representation will be of the form
$\mathcal{C}=\langle\omega p(x)+q'(x),r(x)\rangle$ where $q'(x)\equiv q(x)\,(\mathrm{mod}\,r(x))$.
The size of $\mathcal{C}$ is given by $k=2n-\deg(p)-\deg(r)$, with
a generator matrix consisting of $n-\deg(p)$ cyclic shifts of the
codeword corresponding to $\omega p(x)+q(x)$ and $n-\deg(r)$ cyclic
shifts of the codeword corresponding to $r(x)$. $\mathcal{C}$ is
self-orthogonal (with respect to the trace inner product) if and only
if
\begin{equation}
p(x)r(x^{n-1})\equiv p(x^{n-1})r(x)\equiv0\,(\mathrm{mod}\,x^{n}-1),\label{eq:orthog eqn}
\end{equation}
\begin{equation}
p(x)q(x^{n-1})r(x)\equiv p(x^{n-1})q(x)\,(\mathrm{mod}\,x^{n}-1).\label{eq:orthog eqn 2}
\end{equation}
It is possible to enumerate all the self-orthogonal $(n,2^{k})_{4}$
additive cyclic codes through a slight modification of the method
presented in Ref. \citep{luo2004good}: $r(x)$ ranges over all factors
of $x^{n}-1$; for each $r(x)$, $p(x)$ ranges over the factors of
$x^{n}-1$ of degree $2n-k-\deg(r)$ that satisfy Eq. (\ref{eq:orthog eqn});
and for each pair of $r(x)$ and $p(x)$, $q(x)$ ranges over the
polynomials with $\deg(q)\leq\deg(r)$ that satisfy both Eq. (\ref{eq:orthog eqn 2})
and $q(x)(x^{n}-1)\equiv0\,(\mathrm{mod}\,p(x)r(x))$.

While every additive cyclic code has a ``canonical'' representation
involving two generators, many of them can be described using only
one \citep{huffman2007additive,huffman2008additive} (that is, they
have a generating set comprised of cyclic shifts of a single codeword).
This is guaranteed to be the case if $r(x)=x^{n}-1$ or if $p(x)=x^{n}-1$
and $q(x)$ is a multiple of $r(x)$. However, these are not necessary
conditions for a single-generator representation to exist. For example,
there is a $(5,2^{5})_{4}$ code with $p(x)=1+x$, $q(x)=x^{3}$,
and $r(x)=1+x+x^{2}+x^{3}$, which gives a canonical generator matrix
\begin{equation}
G=\left(\begin{array}{ccccc}
\omega & \omega & 0 & 1 & 0\\
0 & \omega & \omega & 0 & 1\\
1 & 0 & \omega & \omega & 0\\
0 & 1 & 0 & \omega & \omega\\
1 & 1 & 1 & 1 & 1
\end{array}\right);
\end{equation}
however, it is also has the generator matrix
\begin{equation}
G'=\left(\begin{array}{ccccc}
\omega & \omega & 0 & 1 & 0\\
0 & \omega & \omega & 0 & 1\\
1 & 0 & \omega & \omega & 0\\
0 & 1 & 0 & \omega & \omega\\
\omega & 0 & 1 & 0 & \omega
\end{array}\right).
\end{equation}
We can express this code compactly as $\mathcal{C}=\langle\omega\omega010,11111\rangle_{\mathrm{cyc}}\equiv\langle\omega\omega010\rangle_{\mathrm{cyc}}$.

\subsection{Quantum channels}

The action of a quantum channel $\Phi$ on a quantum state described
by the density operator $\rho$ is
\begin{equation}
\Phi(\rho)=\sum_{k}A_{k}\rho A_{k}^{\dagger},
\end{equation}
where the $A_{k}$, called Kraus operators, satisfy $\sum_{k}A_{k}^{\dagger}A_{k}=I$
(the identity operator) \citep{kraus1983states}. We are interested
in qubit systems, for which states belong to a two-dimensional Hilbert
space $\mathcal{H}\cong\mathbb{C}^{2}$. Furthermore, we are concerned
with Pauli channels, which are of the form
\begin{equation}
\Phi(\rho)=p_{I}\rho+p_{X}X\rho X+p_{Y}Y\rho Y+p_{Z}Z\rho Z,
\end{equation}
where $p_{I}+p_{X}+p_{Y}+p_{Z}=1$, and in the computational $\{|0\rangle,|1\rangle\}$
basis
\begin{equation}
X=\left(\begin{array}{cc}
0 & 1\\
1 & 0
\end{array}\right),\,Y=\left(\begin{array}{cc}
0 & -i\\
i & 0
\end{array}\right),\,Z=\left(\begin{array}{cc}
1 & 0\\
0 & -1
\end{array}\right).
\end{equation}
The action of this channel can be interpreted as mapping a pure state
$|\phi\rangle$ to $E|\phi\rangle$ where the error $E$ is $I$ with
probability $p_{I}$, $X$ with probability $p_{X}$, $Y$ with probability
$p_{Y}$, or $Z$ with probability $p_{Z}$ \citep{nielsen2002quantum}.
$X$ can be viewed as a bit-flip operator as $X|0\rangle=|1\rangle$
and $X|1\rangle=|0\rangle$. $Z$ can be viewed as a phase flip as
$Z|0\rangle=|0\rangle$ and $Z|1\rangle=-|1\rangle$. $Y=iXZ$ can
be viewed as a combined bit and phase flip. 

The quantum equivalent of the symmetric channel is the depolarizing
channel, for which $p_{I}=1-p$ and $p_{X}=p_{Y}=p_{Z}=p/3$. For
a number of systems of physical interest, phase-flip errors occur
far more frequently than bit-flip errors \citep{evans2007error,ioffe2007asymmetric}.
We focus on two such asymmetric channels in this paper. The first
is the biased $XZ$ channel, for which the $X$ and $Z$ components
of an error $E\propto X^{u}Z^{v}$, where $u,v\in\mathrm{GF}(2)$,
occur independently with probabilities $q_{X}$ and $q_{Z}$, respectively.
It follows from the independence of the error components that $p_{X}=q_{X}(1-q_{Z})$,
$p_{Z}=q_{Z}(1-q_{X})$, and $p_{Y}=q_{X}q_{Z}$. A typical way to
specify an asymmetric channel with two degrees of freedom is through
the total error probability $p=p_{X}+p_{Y}+p_{Z}$ and bias $\eta=p_{Z}/p_{X}$.
Note that while this definition of bias is consistent with Refs. \citep{robertson2017tailored,sarvepalli2009asymmetric},
some authors give alternate definitions; for example, bias is defined
as $p_{Z}/(p_{X}+p_{Y})$ in Ref. \citep{tuckett2018ultrahigh} and
$(p_{Y}+p_{Z})/(p_{X}+p_{Y})$ in Ref. \citep{napp2012optimal}. Ultimately,
the exact nature of the channel parameterization will have no real
impact on our results, which has lead us to select the simplest definition
of bias. The second channel of interest is the combined amplitude
damping (AD) and dephasing channel, which is described by the non-Pauli
Kraus operators
\begin{align}
A_{0} & =\left(\begin{array}{cc}
1 & 0\\
0 & \sqrt{1-\lambda-\gamma}
\end{array}\right),\\
A_{1} & =\left(\begin{array}{cc}
0 & \sqrt{\gamma}\\
0 & 0
\end{array}\right),\\
A_{2} & =\left(\begin{array}{cc}
0 & 0\\
0 & \sqrt{\lambda}
\end{array}\right).
\end{align}
A Pauli approximation of this channel can be obtained through a process
called Pauli twirling \citep{divincenzo2002quantum,emerson2005scalable,dankert2009exact}.
In particular, the approximate channel is \citep{sarvepalli2009asymmetric}
\begin{align}
\Phi_{T}(\rho) & =\frac{1}{4}\sum_{\sigma\in\{I,X,Y,Z\}}\sigma^{\dagger}\Phi(\sigma\rho\sigma^{\dagger})\sigma\\
 & =\frac{2-\gamma+2\sqrt{1-\lambda-\gamma}}{4}\rho+\frac{\gamma}{4}X\rho X+\frac{\gamma}{4}Y\rho Y\nonumber \\
 & +\frac{2-\gamma-2\sqrt{1-\lambda-\gamma}}{4}Z\rho Z.
\end{align}
Again, this channel has two degrees of freedom ($\lambda$ and $\gamma$)
and can therefore be described in terms of the total error probability
$p$ and bias $\eta=p_{Z}/p_{X}$. Note that in the case of $\eta=1$,
$\Phi_{T}$ reduces to the depolarizing channel. For the sake of brevity,
we will simply refer to $\Phi_{T}$ as the AD channel.

The Pauli matrices are Hermitian, unitary, and anticommute with each
other. Furthermore, they form a group 
\begin{equation}
\mathcal{P}_{1}=\{\pm I,\pm iI,\pm X,\pm iX,\text{\ensuremath{\pm Y,\pm iY,\pm Z,\pm iZ}\}=\ensuremath{\langle X,Y,Z\rangle}}
\end{equation}
called the Pauli group. The $n$-qubit Pauli group $\mathcal{P}_{n}$
is comprised of all $n$-fold tensor product combinations of elements
of $\mathcal{P}_{1}$. Note that when writing elements of $\mathcal{P}_{n}$,
the tensor products are often implied; for example, we may write $I\otimes I\otimes X\otimes I\otimes Y\otimes Z\otimes I\otimes I\in\mathcal{P}_{8}$
as $IIXIYZII$. The weight $w(g)$ of some $g\in\mathcal{P}_{n}$
is the number of nonidentity components from which it is comprised.
It follows from the commutation relations of the Pauli matrices that
any two elements of $\mathcal{P}_{n}$ commute if their nonidentity
components differ in an even number of places; otherwise, they anticommute.

As in the classical case, the noise introduced by a quantum channel
can be protected against using a code. In the qubit case, a code is
a subspace $\mathcal{Q}\subseteq(\mathbb{C}^{2})^{\otimes n}$ whose
elements are again called codewords. These codewords are transmitted
across the combined $n$-qubit channel $\Phi^{\otimes n}$, which,
in the Pauli case, maps a codeword $|\phi\rangle$ to $E|\phi\rangle$
where $E\in\mathcal{P}_{n}$. Similar to the classical case of Eq.
(\ref{eq:classical error probability}), if the error components are
independent, then the probability of an error $E=E_{1}\otimes\dots\otimes E_{n}$
occurring (up to phase) is
\begin{equation}
P(E)=\prod_{i=1}^{n}P(E_{i}),\label{eq:error prob}
\end{equation}
where $P(E_{i})$ is the probability of the error $E_{i}$ occurring
(up to phase) on the single qubit channel $\Phi$. The equivalence
of errors up to phase can be addressed more explicitly by instead
considering $\tilde{E}=\{E,-E,iE,-iE\}\in\mathcal{P}_{n}/\{\pm I,\pm iI\}=\tilde{\mathcal{P}}_{n}$.

\subsection{Stabilizer codes\label{subsec:Stabilizer-codes}}

Stabilizer codes are defined by an Abelian subgroup $\mathcal{S}<\mathcal{P}_{n}$,
called the stabilizer, that does not contain $-I$ \citep{gottesman1997stabilizer}.
The code $\mathcal{Q}$ is the space of states that are fixed by every
element $s_{i}\in\mathcal{S}$; that is,
\begin{equation}
\mathcal{Q}=\{|\phi\rangle\in(\mathbb{C}^{2})^{\otimes n}:s_{i}|\phi\rangle=|\phi\rangle\,\forall\,s_{i}\in\mathcal{S}\}.
\end{equation}
The requirement that $-I\notin\mathcal{S}$ means both that no $s\in\mathcal{S}$
can have a phase factor of $\pm i$, and also that if $s\in\mathcal{S}$,
then $-s\notin\mathcal{S}$. If $\mathcal{S}$ is generated by $M=\{M_{1},\dots,M_{m}\}\subset\mathcal{P}_{n}$,
then it is sufficient (and obviously necessary) for $\mathcal{Q}$
to be stabilized by every $M_{i}$. Assuming that the set of generators
is minimal, which will be the case for all codes considered in this
paper, it can be shown that $\dim(\mathcal{Q})=2^{k}$ where $k=n-m$
\citep{nielsen2002quantum}; that is, $\mathcal{Q}$ encodes the state
of a $k$-qubit system. 

Suppose an error $E$ occurs, mapping some codeword $|\phi\rangle\in\mathcal{\mathcal{Q}}$
to $E|\phi\rangle$. A projective measurement of a generator $M_{i}$
will give the result $+1$ if $[E,M_{i}]=EM_{i}-M_{i}E=0$ or $-1$
if $\{E,M_{i}\}=EM_{i}+M_{i}E=0$. These measurement values define
the syndrome $\boldsymbol{z}\in\mathrm{GF}(2)^{n-k}$ with 
\begin{equation}
z_{i}=\begin{cases}
0 & \mathrm{if}\,[E,M_{i}]=0,\\
1 & \mathrm{if}\,\{E,M_{i}\}=0.
\end{cases}
\end{equation}
Defining $\tilde{\mathcal{S}}=\{\tilde{s}=\{s,-s,is,-is\}:s\in\mathcal{S}\}$,
the syndrome resulting from $\tilde{E}\in\tilde{\mathcal{P}}_{n}$
depends only on which coset of $\tilde{\mathcal{P}}_{n}/N(\tilde{\mathcal{S}})$
it belongs to, where $N(\tilde{\mathcal{S}})=\{g\in\tilde{\mathcal{P}}_{n}:g^{-1}\tilde{\mathcal{P}}_{n}g=\tilde{\mathcal{P}}_{n}\}$
is the normalizer of $\tilde{\mathcal{S}}$ in $\tilde{\mathcal{P}}_{n}$;
furthermore, the effect of $\tilde{E}$ on the code depends only on
which coset of $\tilde{\mathcal{P}}_{n}/\tilde{\mathcal{S}}$ it belongs
to \citep{gottesman1997stabilizer}. Note that as $\tilde{\mathcal{S}}\vartriangleleft N(\tilde{\mathcal{S}})$,
the $2^{n-k}$ cosets of $\tilde{\mathcal{P}}_{n}/N(\tilde{\mathcal{S}})$
are each the union of $2^{2k}$ cosets of $\tilde{\mathcal{P}}_{n}/\tilde{\mathcal{S}}$.
In the classical case, the distance $d$ of a linear code is equal
to the weight of the lowest weight error yielding a trivial syndrome
while having a nontrivial effect on the code. This extends to the
quantum case, with the distance $d$ of a stabilizer code being the
weight of the lowest weight element in $N(\tilde{\mathcal{S}})\backslash\tilde{\mathcal{S}}$
\citep{gottesman1997stabilizer}. An $n$-qubit code of dimension
$2^{k}$ and distance $d$ is called an $[[n,k]]$ or $[[n,k,d]]$
code (the double brackets differentiate it from a classical code). 

Given the equivalence of errors up to an element of the stabilizer,
a MAP decoder will determine the most likely coset 
\begin{equation}
\hat{A}_{\boldsymbol{z}}=\underset{A\in\tilde{\mathcal{P}}_{n}/\tilde{\mathcal{S}}}{\mathrm{argmax}}\,P(A|\boldsymbol{z})
\end{equation}
that is consistent with the syndrome measurement. If $\hat{A}_{\boldsymbol{z}}$
has the representative $\tilde{\hat{E}}=\{\hat{E},-\hat{E},i\hat{E},-i\hat{E}\}$,
then the decoder attempts correction by applying $\hat{E}$ to the
channel output. If $\tilde{E}\in\hat{A}_{\boldsymbol{z}}$, and hence
$\tilde{\hat{E}}\tilde{E}\in\tilde{\mathcal{S}}$, then decoding is
successful; otherwise, a decoding error has occurred. It therefore
follows that the FER is
\begin{equation}
F_{\mathrm{MAP}}=1-\sum_{\boldsymbol{z}\in\mathrm{GF}(2)^{n-k}}P(\hat{A}_{\boldsymbol{z}}).\label{eq:map fer}
\end{equation}
 Unfortunately, this decoding problem has been shown to be \#P-complete
\citep{iyer2015hardness}. Furthermore, the simpler decoding problem
of determining the single most likely error
\begin{equation}
\tilde{\hat{E}}_{\boldsymbol{z}}=\underset{\tilde{E}\in\tilde{\mathcal{P}}_{n}}{\mathrm{argmax}}\,P(\tilde{E}|\boldsymbol{z})\label{eq:se decoder}
\end{equation}
corresponding to the observed syndrome is essentially the same as
the classical decoding problem outlined in Sec. \ref{subsec:Classical-codes}
and hence is also NP-complete \citep{hsieh2011np,kuo2012hardness,kuo2013hardnesses}.
The FER for this decoder is
\begin{equation}
F_{\mathrm{MAP-SE}}=1-\sum_{\boldsymbol{z}\in\mathrm{GF}(2)^{n-k}}P(\tilde{\hat{E}}_{\boldsymbol{z}}\tilde{\mathcal{S}}),\label{eq:map-se fer}
\end{equation}
where ``SE'' stands for ``single error.''

Two stabilizers (or the codes they define) are permutation equivalent
if they are equal up to a relabeling of qubits. As in the classical
case, if two stabilizer codes are permutation equivalent, then they
are both $[[n,k,d]]$ codes; furthermore, they will yield the same
FERs (both $F_{\mathrm{MAP}}$ and $F_{\mathrm{MAP-SE}}$) when the
error components are independently and identically distributed, which
is the case for the channels that we consider. Again, while there
are more general notions of quantum code equivalence, we are always
referring to permutation equivalence in this paper.

The links between stabilizer codes and classical codes can be made
more concrete by representing the elements of $\tilde{\mathcal{P}}_{n}$
as elements of $\mathrm{GF}(2)^{2n}$ \citep{calderbank1997quantum,gottesman1997stabilizer}.
This is achieved via the isomorphism
\begin{equation}
X^{u_{1}}Z^{v_{1}}\otimes\dots\otimes X^{u_{n}}Z^{v_{n}}=X^{\boldsymbol{u}}Z^{\boldsymbol{v}}\leftrightarrow(\boldsymbol{u}|\boldsymbol{v}),
\end{equation}
with the product of elements in $\tilde{\mathcal{P}}_{n}$ corresponding
to addition in $\mathrm{GF}(2)^{2n}$. Furthermore, representatives
of elements in $\tilde{\mathcal{P}}_{n}$ commute if the symplectic
inner product of the binary representations is zero, where the symplectic
inner product of $\boldsymbol{a}=(\boldsymbol{u}|\boldsymbol{v}),\boldsymbol{b}=(\boldsymbol{u}'|\boldsymbol{v}')\in\mathrm{GF}(2)^{2n}$
is $\boldsymbol{a}\circ\boldsymbol{b}=\boldsymbol{u}\cdot\boldsymbol{v}'+\boldsymbol{u}'\cdot\boldsymbol{v}$.
Utilizing this isomorphism, the generators of some stabilizer $\mathcal{S}$
can be used to define the rows of an $m\times2n$ binary matrix
\begin{equation}
H=(H_{X}|H_{Z}),\label{eq:gf2 pcm}
\end{equation}
where $H_{X}$ and $H_{Z}$ are each $m\times n$ matrices. Under
this mapping, the requirement that all stabilizer generators commute
becomes 
\begin{equation}
H_{X}H_{Z}^{T}+H_{Z}H_{X}^{T}=0.\label{eq:commutation criteria}
\end{equation}
Conversely, a $[2n,n+k]$ linear binary code $\mathcal{C}$ with a
parity-check matrix $H$ satisfying this constraint can be used to
define a stabilizer $\mathcal{S}$. Technically, this only specifies
$\tilde{\mathcal{S}}$; however, as previously outlined, it is $\tilde{\mathcal{S}}$
that dictates the effect of an error on a stabilizer code, which means
that the $2^{n-k}$ stabilizers corresponding to $\tilde{\mathcal{S}}$
will all have the same error correction properties (the codes corresponding
to each such stabilizer actually form a partition of $(\mathbb{C}^{2})^{\otimes n}$
\citep{gaitan2010quantum,djordjevic2012quantum}). Without loss of
generality, we can therefore map $\tilde{\mathcal{S}}$ to a particular
stabilizer $\mathcal{S}$ by arbitrarily selecting a phase factor
of $+1$ for all the generators. 

A subclass of stabilizer codes are the Calderbank-Shor-Steane (CSS)
codes \citep{calderbank1996good,steane1996multiple}, which have a
binary representation of the form
\begin{equation}
H=\left(\begin{array}{c|c}
\tilde{H}_{X} & 0\\
0 & \tilde{H}_{Z}
\end{array}\right).\label{eq:css pcm}
\end{equation}
For such codes, the commutation condition of Eq. (\ref{eq:commutation criteria})
becomes $\tilde{H}_{Z}\tilde{H}_{X}^{T}=0$, which is satisfied when
$\mathcal{C}_{X}^{\perp}\subseteq\mathcal{C}_{Z}$, where $\mathcal{C}_{X}$
and $\mathcal{C}_{Z}$ are classical codes defined by the parity-check
matrices $\tilde{H}_{X}$ and $\tilde{H}_{Z}$, respectively. If $\mathcal{C}_{X}=\mathcal{C}_{Z}$,
then this reduces to $\mathcal{C}_{X}^{\perp}\subseteq\mathcal{C}_{X}$,
in which case, the CSS code is called dual containing (DC).

As previously mentioned, the decoding problem of Eq. (\ref{eq:se decoder})
is essentially the same as the classical decoding problem. This link
can be made more explicit by expressing errors within the binary framework
using the mapping $E\propto X^{\boldsymbol{e}_{X}}Z^{\boldsymbol{e}_{Z}}\leftrightarrow\boldsymbol{e}=(\boldsymbol{e}_{X}^{T}|\boldsymbol{e}_{Z}^{T})^{T}$
(where $\boldsymbol{e}_{X}$, $\boldsymbol{e}_{Z}$, and $\boldsymbol{e}$
are column vectors for consistency with the classical case). If the
generators of a stabilizer define the parity-check matrix $H$ for
the binary code $\mathcal{C}$, then the syndrome corresponding to
$E$ can be found by taking the symplectic inner product of $\boldsymbol{e}$
with each row of $H$, which can be written compactly as

\begin{equation}
\boldsymbol{z}=H\left(\begin{array}{c}
\boldsymbol{e}_{Z}\\
\boldsymbol{e}_{X}
\end{array}\right)=HP\boldsymbol{e},
\end{equation}
where 
\begin{equation}
P=\left(\begin{array}{cc}
0 & I\\
I & 0
\end{array}\right).
\end{equation}
With this slight modification to classical syndrome calculation, determining
$\tilde{\hat{E}}_{\boldsymbol{z}}$ in Eq. (\ref{eq:commutation criteria})
corresponds precisely to determining $\hat{\boldsymbol{e}}_{\boldsymbol{z}}$
in Eq. (\ref{eq:syndrome decoder}). Note that some authors avoid
this difference in syndrome calculation by using the mapping $E\propto X^{\boldsymbol{e}_{X}}Z^{\boldsymbol{e}_{Z}}\leftrightarrow\boldsymbol{e}=(\boldsymbol{e}_{Z}^{T}|\boldsymbol{e}_{X}^{T})^{T}$
\citep{mackay2004sparse}, which gives $\boldsymbol{z}=H\boldsymbol{e}$
as in the classical case of Eq. (\ref{eq:classical syndrome definition}).
For a CSS code, the syndrome associated with an error $E\propto X^{\boldsymbol{e}_{X}}Z^{\boldsymbol{e}_{Z}}$
is
\begin{equation}
\boldsymbol{z}=\left(\begin{array}{c}
\tilde{H}_{X}\boldsymbol{e}_{Z}\\
\tilde{H}_{Z}\boldsymbol{e}_{X}
\end{array}\right)=\left(\begin{array}{c}
\boldsymbol{z}_{Z}\\
\boldsymbol{z}_{X}
\end{array}\right).
\end{equation}
This allows the $X$ and $Z$ components of the error to be treated
separately. In particular, $\boldsymbol{e}_{Z}$ can be inferred from
$\tilde{H}_{X}\boldsymbol{e}_{Z}=\boldsymbol{z}_{Z}$, while $\boldsymbol{e}_{X}$
can be inferred from $\tilde{H}_{Z}\boldsymbol{e}_{X}=\boldsymbol{z}_{X}$.
However, this approach is only guaranteed to determine the single
most likely error if the $X$ and $Z$ components of $E$ occur independently,
which is the case for the biased $XZ$ channel but not for the AD
channel among others (see Sec. II E of Ref. \citep{PhysRevA.100.012330}
for a more detailed discussion).

Elements of $\tilde{\mathcal{P}}_{n}$ can also be represented as
elements of $\mathrm{GF}(4)^{n}$ according to the isomorphism \citep{calderbank1996quantum,gottesman1997stabilizer}
\begin{equation}
X^{\boldsymbol{u}}Z^{\boldsymbol{v}}\leftrightarrow\boldsymbol{u}+\omega\boldsymbol{v},
\end{equation}
with the product of elements in $\tilde{\mathcal{P}}_{n}$ corresponding
to addition in $\mathrm{GF}(4)^{n}$. Representatives of elements
in $\tilde{\mathcal{P}}_{n}$ commute if the trace inner product {[}see
Eq. (\ref{eq:trace inner prod}){]} of the corresponding elements
of $\mathrm{GF}(4)^{n}$ is zero. Utilizing this isomorphism, any
$(n,2^{n-k})_{4}$ additive $\mathrm{GF}(4)$ code $\mathcal{C}$
that is self-orthogonal with respect to the trace inner product can
be used to define an $[[n,k]]$ stabilizer code (it is for this reason
that stabilizer codes are sometimes called additive codes). Furthermore,
the generators of the stabilizer $\mathcal{S}$ can be associated
with the rows of a generator matrix $G$ for $\mathcal{C}$. We can
describe a stabilizer code based on properties of $\mathcal{C}$;
for example, if $\mathcal{C}$ is linear and/or cyclic, then we will
also call $\mathcal{S}$ (and the code $\mathcal{Q}$ it defines)
linear and/or cyclic.

Similar to the classical case, when designing a stabilizer code for
the depolarizing channel, the complexity of determining its FER can
be avoided by instead using code distance as something of a proxy.
However, for asymmetric channels, distance becomes a less accurate
metric as the probability of an error occurring no longer depends
only on its weight. One approach in this case is to design codes with
different $X$ and $Z$ distances, which are called $[[n,k,d_{X}/d_{Z}]]$
codes. For these so-called asymmetric codes, $d_{X}$ and $d_{Z}$
are the maximal values for which there is no $\tilde{E}\in N(\tilde{\mathcal{S}})\backslash\tilde{\mathcal{S}}$
where $E\propto X^{\boldsymbol{e}_{X}}Z^{\boldsymbol{e}_{Z}}$ and
both $w(\boldsymbol{e}_{X})<d_{X}$ and $w(\boldsymbol{e}_{Z})<d_{Z}$.
Such codes are typically constructed within the CSS framework, where
$d_{X}=w(\mathcal{C}_{Z}\backslash\mathcal{C}_{X}^{\perp})$ and $d_{X}=w(\mathcal{C}_{X}\backslash\mathcal{C}_{Z}^{\perp})$
\citep{steane1996simple}. Outside of the CSS framework, where the
$X$ and $Z$ components of an error cannot be considered separately,
the distances $d_{X}$ and $d_{Z}$ are somewhat less meaningful and
potentially not even unique. For example, the $(7,2^{6})_{4}$ additive
cyclic code $\langle\omega10\omega100\rangle_{\mathrm{cyc}}$ maps
to the $[[7,1,3]]$ cyclic stabilizer code with $\mathcal{S}=\langle XZIZXII\rangle_{\mathrm{cyc}}$,
which can be considered as a $[[7,1,7/1]]$, $[[7,1,1/7]]$, or $[[7,1,2/3]]$
code. Some examples of asymmetric codes (for qubits) can be found
in Refs. \citep{sarvepalli2009asymmetric,aly2008asymmetric,ezerman2013css,wang2010asymmetric,la2013asymmetric,guenda2013symmetric}.

\section{Approximate FER calculation\label{sec:Approximate-decoding}}

In this paper, we wish to construct stabilizer codes that perform
well on asymmetric channels. In particular, we wish to gauge their
performance directly; that is, we wish to accurately determine the
FER exhibited by a MAP decoder as given in Eq. (\ref{eq:map fer}).
As previously noted, determining this error rate is an \#P-complete
problem. In this section, we therefore investigate lower complexity
methods of approximating $F_{\mathrm{MAP}}$ and derive bounds on
the relative error of these approximations.

\subsection{Limited error set\label{subsec:Limited-error-set}}

In most cases, many of the errors in $\tilde{\mathcal{P}}_{n}$ occur
with very low probability. It seems reasonable to assume that ignoring
these low-probability errors will have little effect on the FER calculation
of Eq. (\ref{eq:map fer}). In particular, suppose we only consider
a subset of errors $\mathcal{E}\subset\tilde{\mathcal{P}}_{n}$. We
can calculate an approximate FER using $\mathcal{E}$ by first partitioning
it by syndrome into the sets $B_{1},\dots,B_{r}$, where $r\leq2^{n-k}$.
Each of these $B_{i}$ is then further partitioned by equivalence
up to an element of $\tilde{\mathcal{S}}$ to give the sets $A_{i1},\dots,A_{is}$,
where $s\leq2^{2k}$. The approximate FER is then
\begin{equation}
F_{\mathcal{E}}=1-\sum_{i=1}^{r}\max_{j}P(A_{ij})=1-\sum_{i=1}^{r}P(\hat{A}_{i}),\label{eq:limited fer}
\end{equation}
where
\begin{equation}
\hat{A}_{i}=\underset{A_{ij}\in B_{i}}{\mathrm{argmax}}\,P(A_{ij}).
\end{equation}
Note that if we wish to explicitly associate a stabilizer $\mathcal{S}$
with $F_{\mathcal{E}}$, then we write $F_{\mathcal{E}}^{\mathcal{S}}$.
In the best case, $\mathcal{E}$ will contain every $\hat{A}_{\boldsymbol{z}}$
in its entirety, which gives $\sum_{\boldsymbol{z}}P(\hat{A}_{\boldsymbol{z}})=\sum_{i=1}^{r}P(\hat{A}_{i})$
and hence $F_{\mathcal{E}}=F_{\mathrm{MAP}}$. In the worst case,
$\sum_{i=1}^{r}P(\hat{A}_{i})=\sum_{\boldsymbol{z}}P(\hat{A}_{\boldsymbol{z}})-(1-P(\mathcal{E}))$,
which gives $F_{\mathcal{E}}=F_{\mathrm{MAP}}+(1-P(\mathcal{E}))$.
In general,
\begin{equation}
0\leq F_{\mathcal{E}}-F_{\mathrm{MAP}}\leq1-P(\mathcal{E}),
\end{equation}
which leads to
\begin{align}
\delta_{\mathcal{E}} & =\frac{F_{\mathcal{E}}-F_{\mathrm{MAP}}}{F_{\mathrm{MAP}}}\nonumber \\
 & \leq\frac{1-P(\mathcal{E})}{F_{\mathrm{MAP}}}\nonumber \\
 & \leq\frac{1-P(\mathcal{E})}{F_{\mathcal{E}}-(1-P(\mathcal{E}))}\label{eq:relative fer bound}\\
 & =\Delta_{\mathcal{E}}.
\end{align}
This bound $\Delta_{\mathcal{E}}$ on the relative error $\delta_{\mathcal{E}}$
in the approximate FER calculation is of practical use as it does
not require any knowledge of $F_{\mathrm{MAP}}$. 

There are two desirable attributes of the set $\mathcal{E}\subset\tilde{\mathcal{P}}_{n}$
used to calculate $F_{\mathcal{E}}$. The first of these, which follows
from Eq. (\ref{eq:relative fer bound}), is for $1-P(\mathcal{E})$
to be less than some predetermined value as this affects the accuracy
of $F_{\mathcal{E}}$. The second is for $|\mathcal{E}|$ to be small
as this reduces the complexity of calculating $F_{\mathcal{E}}$.
It is possible to construct such a set without enumerating $\tilde{\mathcal{P}}_{n}$
in its entirety by exploiting the independence of error components,
which means that the probability of an error occurring depends only
on the number of $I$, $X$, $Y$, and $Z$ components it contains.
Explicitly, the probability of some error $\tilde{E}\in\tilde{\mathcal{P}}_{n}$
occurring is
\begin{equation}
P(\tilde{E})=\prod_{\sigma\in\{I,X,Y,Z\}}p_{\sigma}^{n(\sigma)},\label{eq:error prob2}
\end{equation}
where $n(\sigma)$ is the number of tensor components of $E$ that
are equal to $\sigma$ up to phase. Furthermore, the number of errors
in $\tilde{\mathcal{P}}_{n}$ with a given distribution of components
is \citep{brualdi1977introductory}
\begin{equation}
N=\frac{n!}{n(I)!n(X)!n(Y)!n(Z)!}.
\end{equation}
Therefore, to construct $\mathcal{E}$, we first enumerate all of
the possible combinations of $n(I)$, $n(X)$, $n(Y)$, and $n(Z)$
such that $n(I)+n(X)+n(Y)+n(Z)=n$, which is a straightforward variation
of the integer partition problem \citep{andrews2004integer}. These
combinations are sorted in descending order according to their associated
probability as given in Eq. (\ref{eq:error prob2}). In an iterative
process, we then work through this list of combinations, adding the
$N$ distinct errors associated with each one to $\mathcal{E}$ until
we reach the desired value of $1-P(\mathcal{E})$. This construction
has the added benefit of ensuring that $\mathcal{E}$ is permutation
invariant, which guarantees that $F_{\mathcal{E}}$ will be the same
for equivalent codes.

\begin{figure}
\includegraphics[scale=0.55]{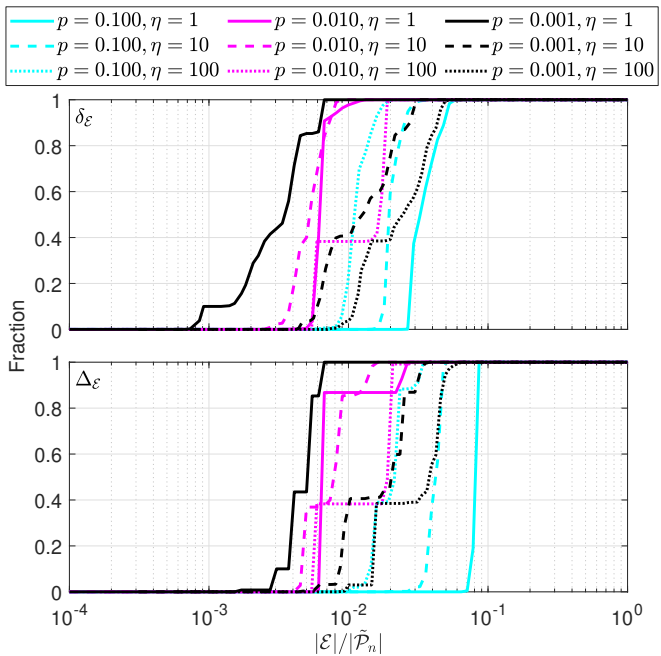}\caption{\label{fig:n7cdf}The fraction of $1\,000$ randomly generated $[[7,1]]$
codes that yield a relative error $\delta_{\mathcal{E}}\protect\leq0.01$
or relative error bound $\Delta_{\mathcal{E}}\protect\leq0.01$ for
varying $|\mathcal{E}|$ and biased $XZ$ channel parameters.}
\end{figure}

\begin{figure}
\includegraphics[scale=0.55]{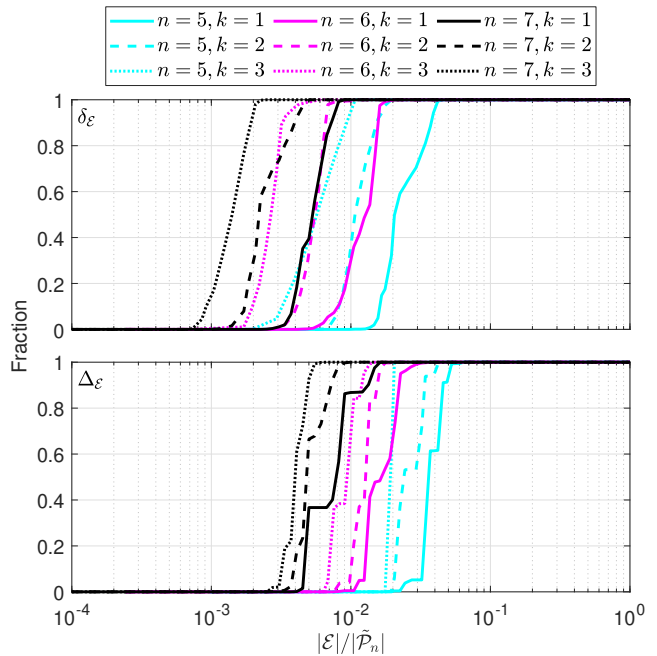}\caption{\label{fig:n5-7k1-3cdf}The fraction of $1\,000$ randomly generated
$[[5\protect\leq n\protect\leq7,1\protect\leq k\protect\leq3]]$ codes
that yield a relative error $\delta_{\mathcal{E}}\protect\leq0.01$
or relative error bound $\Delta_{\mathcal{E}}\protect\leq0.01$ for
a biased $XZ$ channel ($p=0.01$ and $\eta=10$) and varying $|\mathcal{E}|$.}
\end{figure}

For the approximate error rate calculation presented in this section
to be of any real use, it must be accurate even when $\mathcal{E}$
is relatively small. To demonstrate that this is in fact the case,
we have first constructed $1\,000$ random $[[7,1]]$ codes. To produce
a random stabilizer $\mathcal{S}=\langle M_{1},\dots,M_{n-k}\rangle$,
we iteratively select $\tilde{M}_{i}=\{M_{i},-M_{i},iM_{i},-iM_{i}\}$
at random from from $N(\langle\tilde{M}_{1},\dots,\tilde{M}_{i-1}\rangle)\backslash\langle\tilde{M}_{1},\dots,\tilde{M}_{i-1}\rangle$
(note that we arbitrarily use a phase factor $+1$ for each $M_{i}$
as outlined in Sec. \ref{subsec:Stabilizer-codes}). Our only structure
constraint on $\mathcal{S}$ is that it must involve every qubit;
that is, for all $1\leq j\leq n$, there must be some $M_{i}^{(j)}\not\propto I$,
where $M_{i}^{(j)}$ is the $j$th tensor component of $M_{i}$ (if
a stabilizer does not satisfy this constraint, we simply discard it
and construct a new one). For biased $XZ$ channels with $p=0.1$,
$0.01$, or $0.001$ and $\eta=1$, $10$, or $100$, we have then
determined the fraction of the $1\,000$ codes that yield a relative
error $\delta_{\mathcal{E}}\leq0.01$ or relative error bound $\Delta_{\mathcal{E}}\leq0.01$
for varying $|\mathcal{E}|$. The results of this are shown in Fig.
\ref{fig:n7cdf} where it can be seen that, depending on the channel
parameters, only $1-5\%$ of $\tilde{\mathcal{P}}_{n}$ needs to be
considered to yield $\delta_{\mathcal{E}}\leq0.01$ for every code.
As is to be expected, a slightly larger fraction of $\tilde{\mathcal{P}}_{n}$
is required to ensure a relative error bound of $\Delta_{\mathcal{E}}\leq0.01$;
however, in every case this can still be achieved by only considering
between $1-10\%$ of $\tilde{\mathcal{P}}_{n}$. Interestingly, for
higher $p$, increasing $\eta$ reduces the number of errors that
need to be considered, while for lower $p$, this trend is reversed.
Figure \ref{fig:n5-7k1-3cdf} shows the results of a similar analysis
for codes with $5\leq n\leq7$ and $1\leq k\leq3$ on a biased $XZ$
channel with $p=0.01$ and $\eta=10$. It can be seen that increasing
$k$ for fixed $n$ reduces the fraction of errors that must be considered,
which makes sense given that encoding a larger number of qubits will
lead to a higher FER. Furthermore, increasing $n$ for fixed $k$
reduces the fraction of errors that need to be considered, which bodes
well for the analysis of longer codes. We note that changing $p$
and/or $\eta$ has little effect on these observations.

\subsection{Most likely error}

\begin{figure}
\includegraphics[scale=0.55]{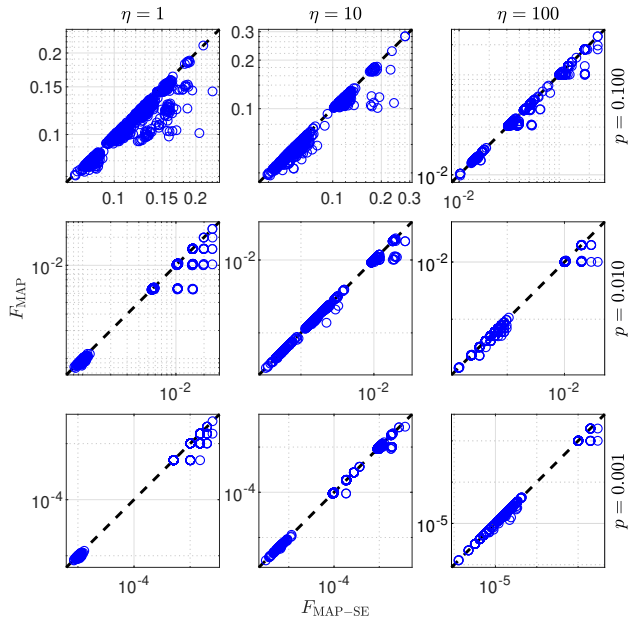}\caption{\label{fig:n7MAPvsMAPSE}$F_{\mathrm{MAP}}$ versus $F_{\mathrm{MAP-SE}}$
for $1\,000$ random $[[7,1]]$ codes on biased $XZ$ channels with
varying parameters. The dotted line gives $F_{\mathrm{MAP}}=F_{\mathrm{MAP-SE}}$.}
\end{figure}

We now consider the decoder of Eq. (\ref{eq:se decoder}) that determines
the single most likely error given a syndrome measurement, which has
an error rate as given in Eq. (\ref{eq:map-se fer}). Note that $F_{\mathrm{MAP-SE}}$
is simpler to calculate than $F_{\mathrm{MAP}}$ as it does not require
a complete partitioning of $\mathcal{\tilde{P}}_{n}$ to form $\tilde{\mathcal{P}}_{n}/\tilde{\mathcal{S}}$.
When using $F_{\mathrm{MAP-SE}}$ as an approximation of $F_{\mathrm{MAP}}$,
the best case scenario is that the most likely coset $\hat{A}_{\boldsymbol{z}}$
will contain $\tilde{\hat{E}}_{\boldsymbol{z}}$ for every $\boldsymbol{z}$,
which gives $F_{\mathrm{MAP-SE}}=F_{\mathrm{MAP}}$. In the worst
case scenario, two things will occur. Firstly, the probability distributions
over every $\hat{A}_{\boldsymbol{z}}$ will be uniform; that is, $P(\hat{A}_{\boldsymbol{z}})/|\tilde{\mathcal{S}}|=P(\hat{A}_{\boldsymbol{z}})/2^{n-k}$
for all $\boldsymbol{z}$. Secondly, the distributions over every
$\tilde{\hat{E}}_{\boldsymbol{z}}\tilde{\mathcal{S}}$ will be sharply
peaked without $P(\tilde{\hat{E}}_{\boldsymbol{z}}\tilde{\mathcal{S}})$
being large; that is, for every $\boldsymbol{z}$, $P(\tilde{\hat{E}}_{\boldsymbol{z}})=P(\hat{A}_{\boldsymbol{z}})/2^{n-k}+\varepsilon$
and $P(\tilde{\hat{E}}_{\boldsymbol{z}}\tilde{\mathcal{S}}\backslash\tilde{\hat{E}}_{\boldsymbol{z}})=\varepsilon'$
for some small $\varepsilon,\varepsilon'\geq0$. In general, it is
therefore the case that
\begin{equation}
F_{\mathrm{MAP}}\leq F_{\mathrm{MAP-SE}}<1-\frac{1-F_{\mathrm{MAP}}}{2^{n-k}}.\label{eq:single error bounds}
\end{equation}
This upper bound on $F_{\mathrm{MAP-SE}}$ is very loose, and in practice,
$F_{\mathrm{MAP-SE}}$ tends to be quite close to $F_{\mathrm{MAP}}$.
To demonstrate this, we have again constructed $1\,000$ random $[[7,1]]$
codes. For each code, we have then determined both $F_{\mathrm{MAP}}$
and $F_{\mathrm{MAP-SE}}$ for the same nine biased $XZ$ channel
parameter combinations considered in Sec. \ref{subsec:Limited-error-set}
($p=0.1$, $0.01$, or $0.001$ and $\eta=1$, $10$, or $100$).
The results of this are shown in Fig. \ref{fig:n7MAPvsMAPSE}. Especially
for the codes yielding a low $F_{\mathrm{MAP}}$, which are the codes
of greatest interest, it can be seen that the difference between $F_{\mathrm{MAP-SE}}$
and $F_{\mathrm{MAP}}$ is often negligible.

$F_{\mathrm{MAP-SE}}$ can itself be approximated using a limited
error set $\mathcal{E}$. We call this approximation $F_{\mathrm{\mathcal{E}-SE}}$,
and it can be calculated in much the same manner as $F_{\mathcal{E}}$.
Again, $\mathcal{E}$ is first partitioned by syndrome to give $B_{1},\dots,B_{r}$.
For each $1\leq i\leq r$, we then determine the most likely error
$\tilde{\hat{E}}_{i}\in B_{i}$, which we use to define $\hat{A}_{i}=\{\tilde{E}\in B_{i}:\tilde{\hat{E}}_{i}\tilde{E}\in\tilde{\mathcal{S}}\}$.
With this altered definition of $\hat{A}_{i}$, $F_{\mathrm{\mathcal{E}-SE}}$
is given by the right-hand side of Eq. (\ref{eq:limited fer}). Furthermore,
the relative error bound of Eq. (\ref{eq:relative fer bound}) also
holds for $F_{\mathrm{\mathcal{E}-SE}}$ with respect to $F_{\mathrm{MAP-SE}}$.
We emphasize that $F_{\mathrm{\mathcal{E}-SE}}$ can be calculated
faster than $F_{\mathrm{\mathcal{E}}}$ as there is no need to fully
partition each $B_{i}$.

\subsection{Most likely error only\label{subsec:Most-likely-error}}

\begin{figure}
\includegraphics[scale=0.55]{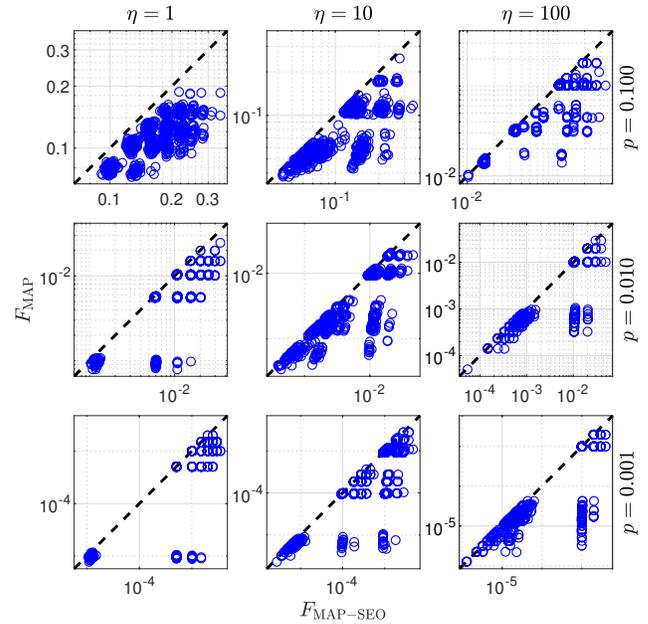}\caption{\label{fig:n7MAPvsMAPSEO}$F_{\mathrm{MAP}}$ versus $F_{\mathrm{MAP-SEO}}$
for $1\,000$ random $[[7,1]]$ codes on biased $XZ$ channels with
varying parameters. The dotted line gives $F_{\mathrm{MAP}}=F_{\mathrm{MAP-SEO}}$.}
\end{figure}

As outlined in Sec. \ref{subsec:Stabilizer-codes}, the single most
likely error decoder for an $[[n,k]]$ stabilizer code can be viewed
as a decoder for an associated $[2n,n+k]$ classical code $\mathcal{C}$.
However, the calculation of $F_{\mathrm{MAP-SE}}$ as in Eq. (\ref{eq:map-se fer})
is more complicated than determining the FER of a classical MAP decoder
as the cosets $\tilde{\hat{E}}_{\boldsymbol{z}}\tilde{\mathcal{S}}$
still need to be enumerated. If we ignore the coset nature of the
error correction, then we get

\begin{equation}
F_{\mathrm{MAP-SEO}}=1-\sum_{\boldsymbol{z}\in\mathrm{GF}(2)^{n-k}}P(\tilde{\hat{E}}_{\boldsymbol{z}}),
\end{equation}
where ``SEO'' stands for ``single error only.'' Note that this
is exactly the FER of the classical decoder for $\mathcal{C}$ as
in Eq. (\ref{eq:classical FER}). Given the nature of the assumptions
leading to Eq. (\ref{eq:single error bounds}), it also holds for
$F_{\mathrm{MAP-SEO}}$. Again, it is a very loose upper bound, and
as can be seen in Fig. \ref{fig:n7MAPvsMAPSEO}, $F_{\mathrm{MAP-SEO}}$
does tend to be somewhat close to $F_{\mathrm{MAP}}$. In particular,
it can be seen that the codes yielding a minimal value of $F_{\mathrm{MAP-SEO}}$
also often yield a near-minimal value of $F_{\mathrm{MAP}}$. 

$F_{\mathrm{MAP-SEO}}$ can also be approximated using a limited error
set to yield $F_{\mathrm{\mathcal{E}-SEO}}$. This involves first
partitioning $\mathcal{E}$ to form $B_{1},\dots,B_{r}$ and then
determining the most likely error $\tilde{\hat{E}}_{i}$ in $B_{i}$.
By defining $\hat{A}_{i}=\tilde{\hat{E}}_{i}$, $F_{\mathrm{\mathcal{E}-SEO}}$
is also given by the right-hand side of Eq. (\ref{eq:limited fer}).
Note that as no partitioning of each $B_{i}$ is required, calculating
$F_{\mathrm{\mathcal{E}-SEO}}$ is less complex than calculating $F_{\mathrm{\mathcal{E}-SE}}$
(or, indeed, $F_{\mathcal{E}}$). The upper bound on relative error
given in Eq. (\ref{eq:relative fer bound}) again holds for $F_{\mathrm{\mathcal{E}-SEO}}$
with respect to $F_{\mathrm{MAP-SEO}}$. Assuming that $\mathcal{E}$
contains the most likely errors in $\tilde{\mathcal{P}}_{n}$, which
is the case for the construction given in Sec. \ref{subsec:Limited-error-set},
we can derive another simple bound. In particular, if $\mathcal{E}$
contains errors corresponding to $r$ different syndromes, then an
error $\tilde{E}'\notin\mathcal{E}$ yielding one of the other $2^{n-k}-r$
possible syndromes must have probability $P(\tilde{E}')\leq\min_{\tilde{E}\in\mathcal{E}}P(\tilde{E})$
(as otherwise it would be an element of $\mathcal{E}$). This gives
\begin{equation}
F_{\mathrm{\mathcal{E}-SEO}}-F_{\mathrm{MAP-SEO}}\leq(2^{n-k}-r)\min_{\tilde{E}\in\mathcal{E}}P(\tilde{E})=\alpha,
\end{equation}
which leads to a combined bound on the relative error of

\begin{align}
\delta_{\mathcal{E}-\mathrm{SEO}} & =\frac{F_{\mathcal{E}-\mathrm{SEO}}-F_{\mathrm{MAP-SEO}}}{F_{\mathrm{MAP-SEO}}}\nonumber \\
 & \leq\frac{\min(1-P(\mathcal{E}),\alpha)}{F_{\mathcal{E}-\mathrm{SEO}}-\min(1-P(\mathcal{E}),\alpha)}\label{eq:combined bound}\\
 & =\Delta_{\mathcal{E}-\mathrm{SEO}}.
\end{align}

\section{Code performance\label{sec:Code-performance}}

In this section, we employ the approximate FER calculation methods
outlined in Sec. \ref{sec:Approximate-decoding} to investigate the
performance of various families of codes on biased $XZ$ and AD channels.
There is a particular focus on the performance of cyclic codes as
it has previously been shown that a $[[7,1,3]]$ cyclic code with
$\mathcal{S}=\langle XZIZXII\rangle_{\mathrm{cyc}}$ performs near-optimally
on the biased $XZ$ channel for a range of error probabilities and
biases \citep{robertson2017tailored}. 

\subsection{$[[7,1]]$ codes\label{subsec:-codes}}

To demonstrate our approach, we first consider the case of $[[7,1]]$
codes. We have constructed all of the $[[7,1]]$ cyclic codes by enumerating
the self-orthogonal additive cyclic $(7,2^{6})_{4}$ codes as outlined
in Sec. \ref{subsec:Cyclic-codes}. There are $11$ such codes, six
of which are inequivalent. Following the lead of Ref. \citep{robertson2017tailored},
we have also constructed $10\,000$ random codes to serve as a point
of comparison. Our random construction, as detailed in Sec. \ref{subsec:Limited-error-set},
differs to that of Ref. \citep{robertson2017tailored} in that we
do not require our codes to have weight-four generators or distance
$d\geq3$. For both biased $XZ$ and AD channels with $p=0.1$, $0.01$,
$0.001$, or $0.0001$ and $\eta=1$, $10$, $100$, or $1\,000$,
we have determined $F_{\mathcal{E}}$ for each code, ensuring that
in every case, $\mathcal{E}$ is large enough to give $\Delta_{\mathcal{E}}\leq0.01$.
This can be achieved without having to construct a new $\mathcal{E}$
for every FER calculation. For some channel type (biased $XZ$ or
AD), channel parameter combination ($p$ and $\eta$ pair), and code
family (random or cyclic), we first construct $\mathcal{E}$, as outlined
in Sec. \ref{subsec:Limited-error-set}, such that $1-P(\mathcal{E})\leq0.1$
and then calculate $F_{\mathcal{E}}$ for every code in the family.
If $\Delta_{\mathcal{E}}>0.01$ for any of these codes, we then add
errors to $\mathcal{E}$ until $1-P(\mathcal{E})\leq0.01$ and recalculate
$F_{\mathcal{E}}$ for these codes. This proceeds iteratively, reducing
$1-P(\mathcal{E})$ by a factor of $10$ each time, until $\Delta_{\mathcal{E}}\leq0.01$
for every code.

\begin{figure}
\includegraphics[scale=0.55]{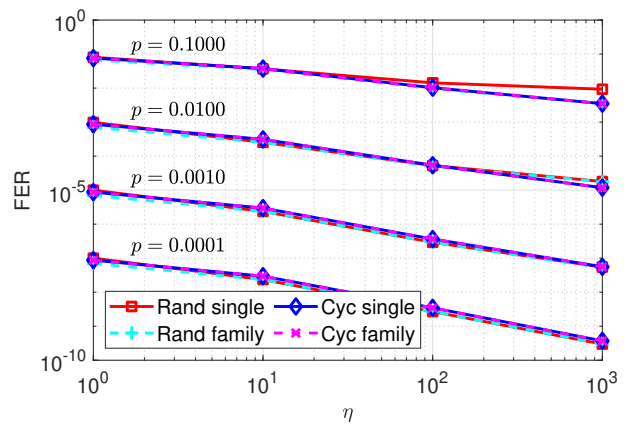}\caption{\label{fig:n7k1BestXZ}FER performance of the best cyclic and random
$[[7,1]]$ codes on biased $XZ$ channels.}
\end{figure}

\begin{figure}
\includegraphics[scale=0.55]{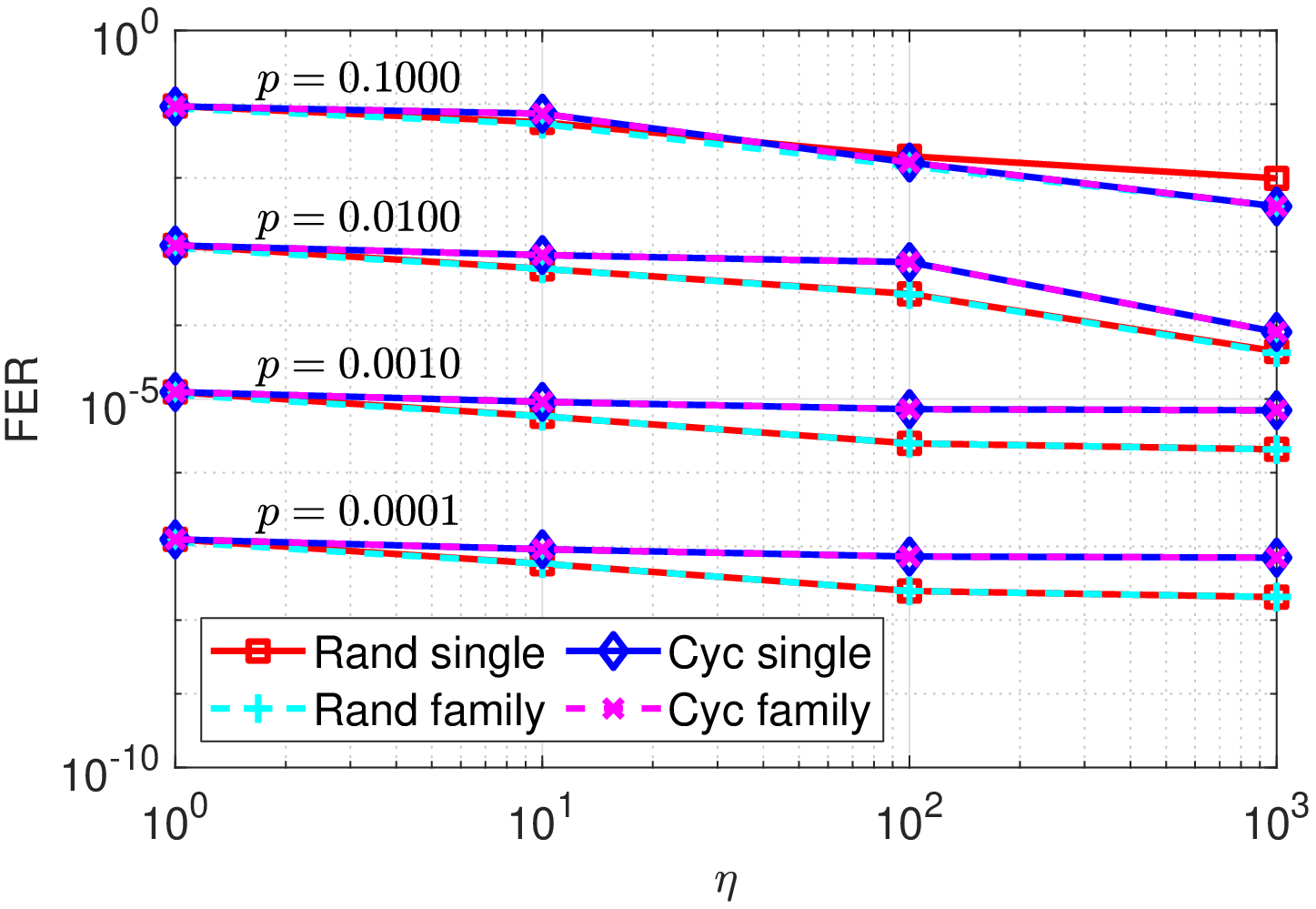}\caption{\label{fig:n7k1BestAD}FER performance of the best cyclic and random
$[[7,1]]$ codes on AD channels.}
\end{figure}

For each channel type, channel parameter combination, and code family,
we report two values. The first of these is simply the lowest FER
of any code in the family, which can be viewed as a performance measure
of the family as a whole. The second is the FER of the code that performs
the best on average across all channel parameter combinations. We
quantify this average performance by taking the geometric mean of
a codes FERs across the associated channels. That is, we take the
best code to be the one with stabilizer
\begin{equation}
\mathcal{S}_{\mathrm{best}}=\underset{\mathcal{\mathcal{S}}\in\mathcal{F}}{\mathrm{argmin}}\,\left(\prod_{i=1}^{N}F_{\mathcal{E}_{i}}^{\mathcal{S}}\right)^{1/N},\label{eq:best code}
\end{equation}
where $\mathcal{F}$ is the family of stabilizers and $\mathcal{E}_{i}$
is the error set associated with one of the $N=16$ channels. Figure
\ref{fig:n7k1BestXZ} shows these values for the biased $XZ$ channel.
It can be seen that for every parameter combination, there is a cyclic
code that performs nearly as well as the best random code. Furthermore,
there is a single cyclic code that performs optimally (among the cyclic
codes) on all channels. In fact, there are three such codes; however,
they are all equivalent to the code with stabilizer $\mathcal{S}=\langle XZIZXII\rangle_{\mathrm{cyc}}$.
The values for the AD channel are shown in Fig. \ref{fig:n7k1BestAD},
where the code with stabilizer $\langle XZIZXII\rangle_{\mathrm{cyc}}$
again performs optimally among the cyclic codes; however, in some
cases, it is outperformed by the best random code by quite a margin,
particularly at lower error probabilities (note that for consistency,
we have used the same random codes for both channel types). At these
low error probabilities, it can also be seen that unlike the biased
$XZ$ channel, increasing the bias does little to decrease the error
rate. Interestingly, the code with stabilizer $\langle YZIZYII\rangle_{\mathrm{cyc}}$,
which is not equivalent to $\langle XZIZXII\rangle_{\mathrm{cyc}}$,
yields the same performance. This is a result of the fact that $p_{X}=p_{Y}$
for the AD channel, which means that applying the permutation $X\leftrightarrow Y$
to a code's stabilizer on any subset of qubits has no effect on its
performance.

Note that the relative error of a geometric mean of FERs, such as
the one in Eq. (\ref{eq:best code}), is bounded by the relative error
of the least accurate individual FER. This follows from
\begin{align}
\left(\prod_{i=1}^{N}F_{\mathcal{E}_{i}}\right)^{1/N} & \leq\left[\prod_{i=1}^{N}(1+\Delta_{\mathcal{E}_{i}})F_{\mathrm{MAP}_{i}}\right]^{1/N}\nonumber \\
 & \leq\max_{i}(1+\Delta_{\mathcal{E}_{i}})\left(\prod_{i=1}^{N}F_{\mathrm{MAP}_{i}}\right)^{1/N},
\end{align}
which gives
\begin{align}
\frac{\left(\prod_{i=1}^{N}F_{\mathcal{E}_{i}}\right)^{1/N}-\left(\prod_{i=1}^{N}F_{\mathrm{MAP}_{i}}\right)^{1/N}}{\left(\prod_{i=1}^{N}F_{\mathrm{MAP}_{i}}\right)^{1/N}} & \leq\max_{i}\Delta_{\mathcal{E}_{i}}.\label{eq:geo mean error bound}
\end{align}

\subsection{Other parameters\label{subsec:Other-parameters}}

We have repeated the analysis of Sec. \ref{subsec:-codes} for codes
with $5\leq n\leq12$ and $1\leq k\leq3$. For each combination of
$n$ and $k$, this has again begun by constructing $10\,000$ random
codes and enumerating the cyclic stabilizer codes. The number of these
cyclic codes is given in the first column of Table \ref{tab:num codes table}.
The first value in each row gives the number of inequivalent codes,
while the value in brackets gives the total number of distinct codes.
Note that for odd $n$, the number of distinct codes we report is
consistent with Ref. \citep{huffman2007additive}. To the best of
our knowledge, neither the number of distinct codes with even $n$
or the number of inequivalent codes with any $n$ has previously been
published (Ref. \citep{huffman2008additive} does give total number
of distinct $[[n,k\leq n]]$ cyclic codes, but it does not include
the number for each specific $k$). Note that in some cases, there
are no cyclic codes.

\begin{table}
\caption{\label{tab:num codes table}The number of inequivalent (distinct)
$[[n,k]]$ cyclic codes, single-generator cyclic codes, cyclic codes
with weight-four generators, cyclic CSS codes, dual-containing CSS
codes, and linear cyclic codes.}
\begin{tabular*}{8.6cm}{@{\extracolsep{\fill}}ccccccc}
\toprule 
$[[n,k]]$ & Cyc & One gen & $w=4$ & CSS & DC CSS & Lin\tabularnewline
\midrule
$[[5,1]]$ & $4\,(5)$ & $4\,(5)$ & $4\,(5)$ & $2\,(2)$ & $0$ & $1\,(2)$\tabularnewline
$[[5,2]]$ & $0\,(0)$ & $0\,(0)$ & $0\,(0)$ & $0\,(0)$ & $0$ & $0\,(0)$\tabularnewline
$[[5,3]]$ & $0\,(0)$ & $0\,(0)$ & $0\,(0)$ & $0\,(0)$ & $0$ & $0\,(0)$\tabularnewline
$[[6,1]]$ & $21\,(21)$ & $18\,(18)$ & $15\,(15)$ & $6\,(6)$ & $0$ & $0\,(0)$\tabularnewline
$[[6,2]]$ & $35\,(42)$ & $30\,(36)$ & $17\,(21)$ & $9\,(9)$ & $2$ & $2\,(3)$\tabularnewline
$[[6,3]]$ & $12\,(15)$ & $12\,(15)$ & $3\,(6)$ & $4\,(4)$ & $0$ & $0\,(0)$\tabularnewline
$[[7,1]]$ & $6\,(11)$ & $5\,(9)$ & $6\,(11)$ & $3\,(4)$ & $1$ & $1\,(2)$\tabularnewline
$[[7,2]]$ & $0\,(0)$ & $0\,(0)$ & $0\,(0)$ & $0\,(0)$ & $0$ & $0\,(0)$\tabularnewline
$[[7,3]]$ & $15\,(54)$ & $15\,(54)$ & $0\,(0)$ & $4\,(8)$ & $0$ & $0\,(0)$\tabularnewline
$[[8,1]]$ & $57\,(87)$ & $30\,(48)$ & $24\,(33)$ & $8\,(8)$ & $0$ & $0\,(0)$\tabularnewline
$[[8,2]]$ & $46\,(79)$ & $27\,(48)$ & $19\,(25)$ & $7\,(7)$ & $3$ & $1\,(1)$\tabularnewline
$[[8,3]]$ & $33\,(63)$ & $21\,(48)$ & $12\,(15)$ & $6\,(6)$ & $0$ & $0\,(0)$\tabularnewline
$[[9,1]]$ & $15\,(27)$ & $15\,(27)$ & $9\,(21)$ & $4\,(4)$ & $1$ & $0\,(0)$\tabularnewline
$[[9,2]]$ & $15\,(27)$ & $15\,(27)$ & $0\,(0)$ & $4\,(4)$ & $0$ & $0\,(0)$\tabularnewline
$[[9,3]]$ & $5\,(9)$ & $5\,(9)$ & $3\,(3)$ & $2\,(2)$ & $1$ & $0\,(0)$\tabularnewline
$[[10,1]]$ & $42\,(63)$ & $39\,(60)$ & $21\,(33)$ & $6\,(6)$ & $0$ & $0\,(0)$\tabularnewline
$[[10,2]]$ & $14\,(21)$ & $13\,(20)$ & $11\,(15)$ & $3\,(3)$ & $6$ & $2\,(3)$\tabularnewline
$[[10,3]]$ & $0\,(0)$ & $0\,(0)$ & $0\,(0)$ & $0\,(0)$ & $0$ & $0\,(0)$\tabularnewline
$[[11,1]]$ & $9\,(33)$ & $9\,(33)$ & $9\,(33)$ & $2\,(2)$ & $2$ & $0\,(0)$\tabularnewline
$[[11,2]]$ & $0\,(0)$ & $0\,(0)$ & $0\,(0)$ & $0\,(0)$ & $0$ & $0\,(0)$\tabularnewline
$[[11,3]]$ & $0\,(0)$ & $0\,(0)$ & $0\,(0)$ & $0\,(0)$ & $3$ & $0\,(0)$\tabularnewline
$[[12,1]]$ & $300\,(465)$ & $162\,(288)$ & $51\,(75)$ & $20\,(20)$ & $0$ & $0\,(0)$\tabularnewline
$[[12,2]]$ & $536\,(768)$ & $288\,(432)$ & $65\,(81)$ & $35\,(35)$ & $11$ & $2\,(3)$\tabularnewline
$[[12,3]]$ & $312\,(528)$ & $198\,(360)$ & $27\,(30)$ & $26\,(26)$ & $0$ & $0\,(0)$\tabularnewline
\bottomrule
\end{tabular*}
\end{table}

\begin{figure}
\includegraphics[scale=0.55]{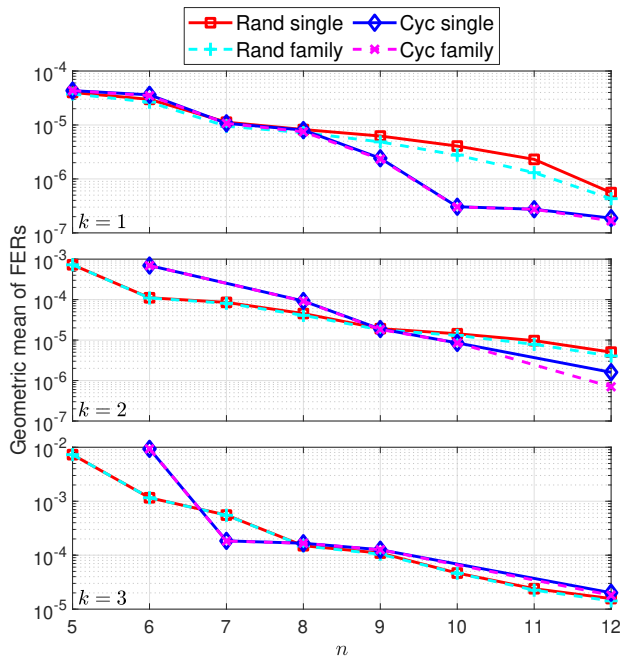}

\caption{\label{fig:n10k3geoMean}The geometric mean of FERs for codes on biased
$XZ$ channels with $p=0.1$, $0.01$, $0.001$, or $0.0001$ and
$\eta=1$, $10$, $100$, or $1\,000$.}

\end{figure}

\begin{figure}
\includegraphics[scale=0.55]{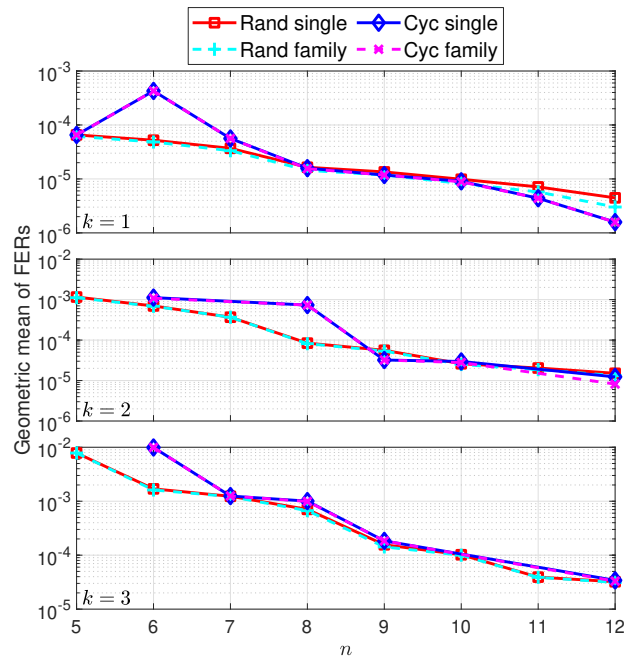}\caption{\label{fig:n10k3geoMean-1}The geometric mean of FERs for codes on
AD channels with $p=0.1$, $0.01$, $0.001$, or $0.0001$ and $\eta=1$,
$10$, $100$, or $1\,000$.}
\end{figure}

For each channel type, code family, and pair of $n$ and $k$, we
report two values. The first of these is the geometric mean of the
FERs for the single best code as defined in Eq. (\ref{eq:best code});
that is, 

\begin{equation}
\lambda=\min_{\mathcal{\mathcal{S}}\in\mathcal{F}}\left(\prod_{i=1}^{N}F_{\mathcal{E}_{i}}^{\mathcal{S}}\right)^{1/N}.
\end{equation}
The second value is the geometric mean of the minimum FERs of all
codes in a family for each channel; that is
\begin{equation}
\mu=\left(\prod_{i=1}^{N}\min_{\mathcal{S}\in\mathcal{F}}F_{\mathcal{E}_{i}}^{\mathcal{S}}\right)^{1/N},
\end{equation}
which can again be viewed as a performance measure of the family as
a whole. Figure \ref{fig:n10k3geoMean} shows these values for the
biased $XZ$ channel. It can be seen that for both the random and
cyclic codes, there is typically a single code that performs nearly
as well as the family as a whole across the $16$ different channels
considered. Furthermore, when $[[n,k]]$ cyclic codes exist, there
is often one that performs as well as or better than the best random
code we have created. In fact, for $n\geq9$ and $k=1$, the best
cyclic codes significantly outperform the best random codes. The results
for the AD channel are given in Fig. \ref{fig:n10k3geoMean-1}. Again,
where $[[n,k]]$ cyclic codes exist, they typically perform favorably
compared to the random codes. However, any performance advantages
over the random codes are less pronounced than in the biased $XZ$
case.

Generators for the best cyclic codes on both the biased $XZ$ and
AD channels can be found in Table \ref{tab:best cyc cocdes} (for
reference, we also their distances). In particular, we list generators
for all codes that yield a geometric mean of FERs within $1\%$ of
the minimum value we have observed (these are all codes that could
conceivably be optimal within our margin of error). There are a few
notable properties of these codes. The first of these is that they
can all be expressed using a single generator. While, as shown in
the second column of Table \ref{tab:num codes table}, a large number
of codes have such a representation, this is still a somewhat surprising
result. It can also be seen that in nearly every case, there are codes
that perform well for both the biased $XZ$ and AD channels (the only
exceptions to this are the $[[6,1]]$, $[[6,2]]$, and $[[10,2]]$
cases). A third property of note is that the codes for the AD channel
typically come in pairs, one being an $X\leftrightarrow Y$ permuted
version of the other. This is to be expected given the partial channel
symmetry outlined in Sec. \ref{subsec:-codes}. The only two exceptions
to this are the $[[5,1]]$ and $[[10,2]]$ cases, where the single
code given is invariant under an $X\leftrightarrow Y$ permutation
(up to a permutation of qubit labels).

\subsection{Hill climbing\label{subsec:Hill-climbing}}

The results of Sec. \ref{subsec:Other-parameters}, particularly those
for $[[n\geq9,1]]$ codes on the biased $XZ$ channel, show that constructing
$10\,000$ random codes is not a reliable way of finding a good code
for larger $n$. One approach to find better codes would be to simply
increase the size of the random search. However, even with the reduction
in error set size afforded by the approach of Sec. \ref{subsec:Limited-error-set},
this quickly becomes computationally impractical. As such, we need
a more efficient search strategy. To achieve this, we use the observation
of Sec. \ref{subsec:Most-likely-error} that codes yielding a low
$F_{\mathcal{E}-\mathrm{SEO}}$ tend to also yield a low $F_{\mathcal{E}}$
(recall that $F_{\mathcal{E}}\leq F_{\mathcal{E-\mathrm{SEO}}}$).
We can therefore reduce the search to finding codes that yield a low
$F_{\mathcal{E}-\mathrm{SEO}}$, which is beneficial as it is typically
several orders of magnitude faster to calculate $F_{\mathcal{E}-\mathrm{SEO}}$
than it is to calculate $F_{\mathcal{E}}$ to the same accuracy. 

We start by considering the problem of finding codes that perform
well for a single channel parameter combination. That is, we want
to find a stabilizer $\mathcal{S}$ that yields a low $F_{\mathcal{E}-\mathrm{SEO}}^{\mathcal{S}}$.
We have found a simple hill climbing search strategy to be effective
at this. This involves first constructing $\mathcal{S}$ at random.
$\mathcal{S}$ is then mutated (modified) somehow to produce $\mathcal{S}'$,
and if $F_{\mathcal{E}-\mathrm{SEO}}^{\mathcal{S}'}\leq F_{\mathcal{E}-\mathrm{SEO}}^{\mathcal{S}}$,
then $\mathcal{S}$ is replaced with $\mathcal{S}'$. This process
repeats for a predetermined number of iterations, after which we calculate
$F_{\mathcal{E}}^{\mathcal{S}}$ to quantify the actual performance
of the code. Similar to the random search outlined in Sec. \ref{subsec:-codes},
we ensure that the relative error of all approximate FER calculations
is less than $1\%$. To achieve this, we again initially construct
$\mathcal{E}$ such that $1-P(\mathcal{E})\leq0.1$, and if $\Delta_{\mathcal{E}-\mathrm{SEO}}>0.01$
($\Delta_{\mathcal{E}}>0.01$) for any calculation of $F_{\mathcal{E}-\mathrm{SEO}}$
($F_{\mathcal{E}}$), then we add errors to $\mathcal{E}$ to reduce
$1-P(\mathcal{E})$ by a factor of $10$ and recalculate the error
rate. To better explore the space of possible stabilizers, we run
a number of these hill climbing instances in parallel (this is often
called hill climbing with random restarts \citep{lones2011sean}).

\begin{figure}
\includegraphics[scale=0.55]{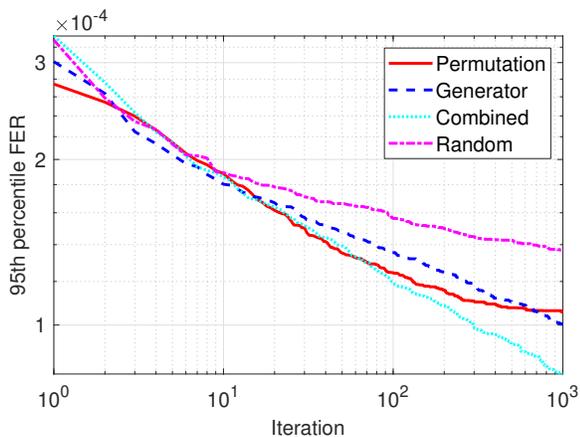}\caption{\label{fig:mut comparison}95th percentile $F_{\mathcal{E}-\mathrm{SEO}}$
found by $1\,000$ hill climbing instances based on various mutation
methods for $[[9,1]]$ codes on a biased $XZ$ channel ($p=0.01$
and $\eta=10$).}

\end{figure}

The choice of a mutation operator that maps $\mathcal{S}$ to $\mathcal{S}'$
is limited by the requirement that $\mathcal{S}'$ must be a stabilizer.
We consider two types of mutation that satisfy this constraint. The
first of these involves permuting the nonidentity Pauli matrices of
all stabilizer elements at any given index $1\leq i\leq n$ with probability
$1/n$. Note that these permutations correspond to a multiplication
of coordinates of the associated classical $\mathrm{GF}(4)$ code
by a nonzero scalar $\alpha\in\mathrm{GF}(4)$ followed by a possible
conjugation. The second mutation method involves first removing any
given generator $M_{i}$ of $\mathcal{S}=\langle M_{1},\dots,M_{n-k}\rangle$
with probability $1/(n-k)$ and then adding generators, as outlined
in Sec. \ref{subsec:Limited-error-set}, to form $\mathcal{S}'$.
When performing this generator mutation, we still require that all
qubits are involved in the stabilizer; if this is not achieved after
adding the new generators, we remove them and try again. To compare
these two mutation operators, we consider $[[9,1]]$ codes on the
biased $XZ$ channel with $p=0.01$ and $\eta=10$. We have run $1\,000$
hill climbing instances, each for a maximum of $1\,000$ iterations.
Across all of these instances, Fig. \ref{fig:mut comparison} shows
the 95th percentile $F_{\mathcal{E}-\mathrm{SEO}}$ at each iteration;
that is, it shows the $50$th lowest $F_{\mathcal{E}-\mathrm{SEO}}$
(we have chosen to show this value as it reflects the performance
of the best codes while having less potential variance than showing
the FER of the single best code). As a control, we have also tested
random mutation, which involves simply creating $\mathcal{S}'$ at
random (this reduces hill climbing to a random search). It can be
seen that both the permutation and generator mutation outperform this
random mutation, with the permutation mutation performing best initially
but then tapering off somewhat. Finally, we have tested a combination
of the two mutation methods (a generator mutation followed by a permutation
mutation), which can be seen to perform better than either of the
methods individually. 

\subsection{Multiobjective hill climbing\label{subsec:Multi-objective-hill-climbing}}

The results of Sec. \ref{subsec:Other-parameters} suggest that there
are typically codes that perform well across a range of channel parameter
combinations. We can search for such codes by building on the hill
climbing algorithm outlined in Sec. \ref{subsec:Hill-climbing}. In
particular, instead of comparing $F_{\mathcal{E}-\mathrm{SEO}}^{\mathcal{S}'}$
to $F_{\mathcal{E}-\mathrm{SEO}}^{\mathcal{S}}$, we compute and compare
the geometric means $(\prod_{i=1}^{N}F_{\mathcal{E}_{i}-\mathrm{SEO}}^{\mathcal{S}'})^{1/N}$
and $(\prod_{i=1}^{N}F_{\mathcal{E}_{i}-\mathrm{SEO}}^{\mathcal{S}})^{1/N}$
of the FERs for $N$ channel parameter combinations. Following Eq.
(\ref{eq:geo mean error bound}), we ensure that these geometric means
are accurate to within $1\%$ by keeping each of the individual $\Delta_{\mathcal{E}_{i}-\mathrm{SEO}}\leq0.01$
as outlined in Sec. \ref{subsec:Hill-climbing}. Again, we run a number
of these hill climbing instances in parallel, and at the end of each
one, we calculate $(\prod_{i=1}^{N}F_{\mathcal{E}_{i}}^{\mathcal{S}})^{1/N}$.
Note that for $N=1$, this search reduces to that of Sec. \ref{subsec:Hill-climbing}.

\begin{figure}
\includegraphics[scale=0.55]{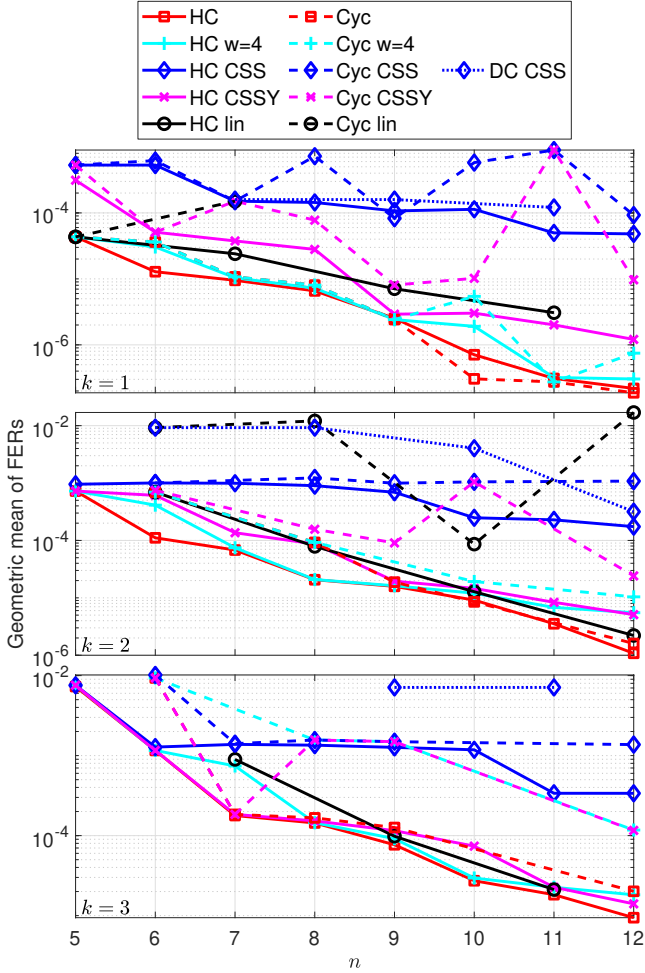}\caption{\label{fig:XZ HC FERs}The performance (geometric mean of FERs) of
the best $[[5\protect\leq n\protect\leq12,1\protect\leq k\protect\leq3]]$
codes found via hill climbing for biased $XZ$ channels with with
$p=0.1$, $0.01$, $0.001$, or $0.0001$ and $\eta=1$, $10$, $100$,
or $1\,000$. Also shown is the performance of the best cyclic codes
and dual-containing CSS codes.}
\end{figure}

\begin{figure}
\includegraphics[scale=0.55]{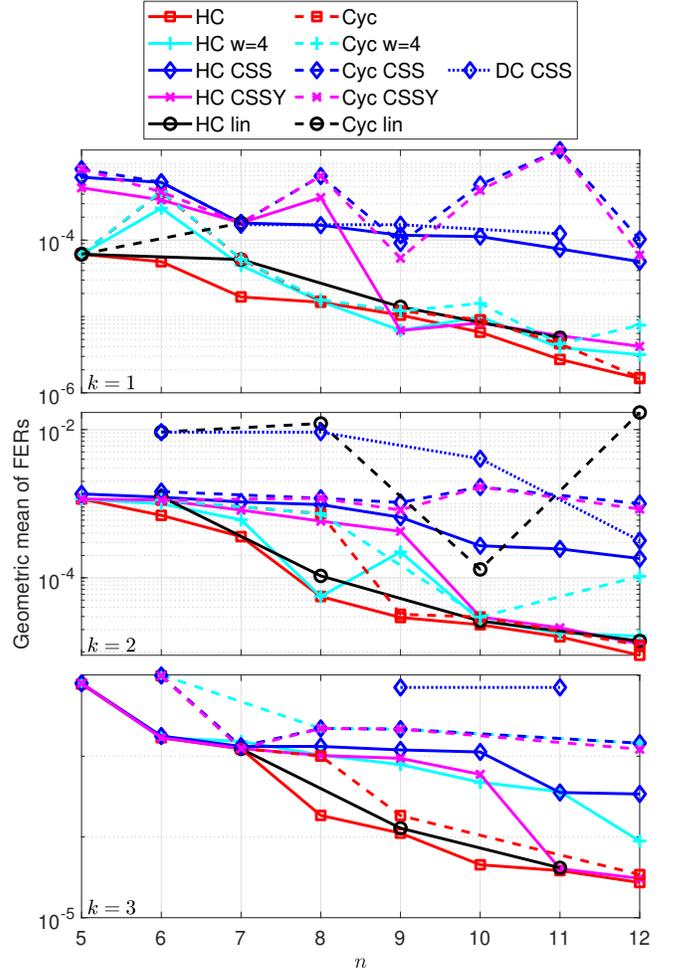}\caption{\label{fig:AD HC FERs}The performance (geometric mean of FERs) of
the best $[[5\protect\leq n\protect\leq12,1\protect\leq k\protect\leq3]]$
codes found via hill climbing for AD channels with with $p=0.1$,
$0.01$, $0.001$, or $0.0001$ and $\eta=1$, $10$, $100$, or $1\,000$.
Also shown is the performance of the best cyclic codes and dual-containing
CSS codes.}
\end{figure}

We have performed such searches for the same cases considered in Sec.
\ref{subsec:Other-parameters} (that is, codes with $5\leq n\leq12$
and $1\leq k\leq3$ for biased $XZ$ and AD channels with $p=0.1$,
$0.01$, $0.001$, or $0.0001$ and $\eta=1$, $10$, $100$, or $1\,000$).
For each combination of $n$, $k$, and channel type, we have run
$1\,000$ hill climbing instances based on the combined generator
and permutation mutation, each for $1\,000$ iterations. Figure \ref{fig:XZ HC FERs}
compares the performance (that is, the geometric mean of FERs) of
the best codes found in this way to that of the best cyclic codes
(the other values shown will be detailed in Secs. \ref{subsec:Weight-four-codes}
to \ref{subsec:Linear-codes}). It can be seen that in all but the
$[[10,1]]$ case, the best code found via hill climbing is either
as good as or better than the best cyclic code. Very similar results
can be seen in Fig. \ref{fig:AD HC FERs} for the AD channel, where
the best code found via hill climbing performs as well as or better
than the best cyclic code in every instance. Generators for the best
codes we have found for the biased $XZ$ and AD channels can be found
in Tables \ref{tab:XZ HC codes} and \ref{tab:AD HC codes}, respectively.

\subsection{Weight-four codes\label{subsec:Weight-four-codes}}

Through slight modification of the hill climbing algorithm, we can
search for good codes that satisfy structure constraints. The first
constraint we consider is the requirement that the stabilizer has
a representation involving only weight-four generators; such codes
are of practical interest as their syndrome measurements involve fewer
qubits, and are hence less complex, than those for codes with high-weight
generators. The first modification required to search for these codes,
which is somewhat obvious, is to ensure the initial random stabilizer
has weight-four generators. This also extends to the generator permutation;
that is, any generator added to replace a removed one must also have
weight four. No change to the permutation mutation is required as
it preserves the weight of stabilizer elements. We compare the codes
found via this constrained hill climbing search to the cyclic codes
with a weight-four generator representation. The number of such cyclic
codes is given in the third column of Table \ref{tab:num codes table},
where it can be seen that they are reasonably plentiful. 

The performance of the weight-four codes found via hill climbing for
the biased $XZ$ channel is shown in Fig. \ref{fig:XZ HC FERs}. It
can be seen that in a lot of cases, these codes perform nearly as
well as those found using unconstrained hill climbing in Sec. \ref{subsec:Multi-objective-hill-climbing}.
The performance of the weight-four cyclic codes is more varied. In
some cases, they are optimal (among the cyclic codes), while in others,
they perform relatively poorly. Figure \ref{fig:AD HC FERs} shows
that the performance of the weight-four codes found via hill climbing
for the AD channel is somewhat mixed, ranging from outperforming the
unconstrained $[[9,1]]$ codes to performing very poorly for $k=3$
and $n\geq8$. The performance of the weight-four cyclic codes relative
to the best unconstrained cyclic codes is much the same as for the
biased $XZ$ channel. Generators for the best weight-four codes found
via hill climbing can be found in Tables \ref{tab:XZ HC codes w4}
and \ref{tab:AD HC codes w4}, and generators for the best cyclic
codes are given in Table \ref{tab:best w4 cyc cocdes}.

\subsection{CSS codes}

We next consider CSS codes, which, as outlined in Sec. \ref{subsec:Stabilizer-codes},
are codes that can be represented using generators that contain either
only $X$ or only $Z$ matrices as their nonidentity elements. Similar
to the search for weight-four codes, we must modify both the initial
stabilizer construction and the generator permutation. In particular,
when adding a new generator, we will select a suitable $X$-only element
half the time and a $Z$-only element the other half. Another required
modification is the removal of the permutation mutation as, in general,
it does not map CSS codes to CSS codes. We also consider cyclic CSS
codes, which can be thought of in two equivalent ways. They can be
viewed as codes with a binary representation where $\tilde{H}_{X}$
and $\tilde{H}_{Z}$ each correspond to a binary cyclic code. Alternatively,
they can be considered in the $\mathrm{GF}(4)$ framework as additive
cyclic codes that can be represented by an $1$-only cyclic generator
and/or an $\omega$-only cyclic generator. The number of these cyclic
CSS codes is given in column four of Table \ref{tab:num codes table}.
We also consider the family of dual-containing CSS codes to generalize
the result of Ref. \citep{robertson2017tailored}, where it was shown
that the $[[7,1,3]]$ Steane code \citep{steane1996error}, which
has 
\begin{equation}
\tilde{H}_{X}=\tilde{H}_{Z}=\left(\begin{array}{ccccccc}
1 & 0 & 1 & 0 & 1 & 0 & 1\\
0 & 1 & 1 & 0 & 0 & 1 & 1\\
0 & 0 & 0 & 1 & 1 & 1 & 1
\end{array}\right),
\end{equation}
 performs poorly on the biased $XZ$ channel. We have constructed
these codes by enumerating all of the inequivalent binary self-orthogonal
codes using \textsc{sagemath} \citep{sage} (recall that a generator
matrix for a binary-self orthogonal code is the parity-check matrix
for a dual-containing code). The number of such codes is given in
column five of Table \ref{tab:num codes table}. Note that there can
only be an $[[n,k]]$ dual-containing CSS code if $n-k$ is even;
furthermore, even when $n-k$ is even, not many of them exist for
the parameters considered. 

As can be seen for the biased $XZ$ channel in Fig. \ref{fig:XZ HC FERs},
both the CSS codes found via hill climbing and the cyclic CSS codes
perform poorly compared to their non-CSS counterparts. This performance
can be improved by following the modification outlined in Ref. \citep{tuckett2018ultrahigh},
which involves applying the permutation $Z\leftrightarrow Y$ to the
code's generators (this is motivated by the fact that $Z$-only generators
commute with any $Z$-only error, meaning that they often provide
no information about an error when $\eta$ is large). Given the nature
of this modification, we call such codes CSSY codes. We have performed
a hill climbing search for CSSY codes, and it can be seen that they
perform significantly better than the standard CSS codes; however,
they are still outperformed by non-CSS codes in most instances. Similarly,
while the cyclic CSSY codes perform better than the cyclic CSS codes,
there is often a significant performance gap to the non-CSS cyclic
codes. The dual-containing CSS codes perform poorly across the board,
which can at least partially be attributed to the fact that they must
have $d_{X}=d_{Z}$. Furthermore, their performance cannot be improved
as they are invariant under a $Z\leftrightarrow Y$ permutation. As
shown in Fig. \ref{fig:AD HC FERs}, the results on the AD channel
are similar to those for the biased $XZ$ channel. Both the CSS codes
found via hill climbing and the cyclic CSS codes perform poorly compared
to the non-CSS codes. In this case, the performance gain of the CSSY
codes over the CSS codes is less pronounced. A notable exception to
this is the $[[9,1]]$ case where the best CSSY code found via hill
climbing outperforms the best unrestricted code found. Somewhat surprisingly,
after applying an $X\leftrightarrow Y$ permutation to the second,
fourth, fifth, sixth, and ninth qubits, this code is equivalent to
the best code with weight-four generators found in Sec. \ref{subsec:Weight-four-codes}.
Again, the performance of the dual-containing CSS codes is very poor
compared to nearly all other codes considered. Generators for the
best CSSY codes found via hill climbing can be found in Tables \ref{tab:XZ HC codes CSSY}
and \ref{tab:AD HC codes CSSY}. We omit the standard CSS codes found
via hill climbing and the cyclic CSS(Y) codes due to their poor performance.

\subsection{Linear codes\label{subsec:Linear-codes}}

The dual-containing CSS codes considered in the previous section are
examples of linear stabilizer codes. An additive $(n,2^{n-k})_{4}$
code $\mathcal{C}$ is linear if and only if it has a generating set
of the form $\mathcal{B}=\{\boldsymbol{b}_{1},\dots,\boldsymbol{b}_{(n-k)/2},\omega\boldsymbol{b}_{1},\dots,\omega\boldsymbol{b}_{(n-k)/2}\}$.
This corresponds to the stabilizer having generators of the form $\mathcal{S}=\langle M_{1},\dots,M_{(n-k)/2},\bar{M}_{1},\dots,\bar{M}_{(n-k)/2}\rangle$,
where $\bar{M}_{i}$ is a version of $M_{i}$ that has been subjected
to the permutation $(X,Y,Z)\rightarrow(Z,X,Y)$. To search for such
codes, we must first modify the initial construction and generator
mutations. In particular, we add or remove the generators $M_{i}$
and $\bar{M}_{i}$ in pairs. To preserve linearity, the permutation
mutation has to be restricted to permutations corresponding to a multiplication
of a coordinate of $\mathcal{C}$ by $\omega$ or $\bar{\omega}$.
That is, the permutation must either be $(X,Y,Z)\rightarrow(Z,X,Y)$
or $(X,Y,Z)\rightarrow(Y,Z,X)$. We also consider linear cyclic codes,
the structure of which is outlined in Sec. \ref{subsec:Cyclic-codes}.
The number of such codes is given in column six of Table \ref{tab:num codes table}.
Like the dual-containing CSS codes, $[[n,k]]$ linear codes can only
exist for even $n-k$; furthermore, while $n-k$ is even for $[[5,3]]$
codes, there are no linear codes with these parameters that involve
every qubit.

As shown in Fig. \ref{fig:XZ HC FERs}, the linear codes found via
hill climbing perform reasonably well on the biased $XZ$ channel.
The performance of the linear cyclic codes is somewhat less impressive,
with there being a significant gap in performance to the more general
additive cyclic codes. This can potentially be attributed to the fact
that, at least for the code parameters considered, there are very
few linear codes. As can be seen in Fig. \ref{fig:AD HC FERs}, the
linear codes found via hill climbing for the AD channel perform better
than those on the biased $XZ$ channel, particularly in the $k=3$
case. However, the linear cyclic codes still perform poorly. The best
linear codes found via hill climbing are given in Tables \ref{tab:XZ HC codes lin}
and \ref{tab:AD HC codes lin}. We omit the linear cyclic codes due
to their poor performance.

\section{Conclusion\label{sec:Conclusion}}

We have shown that the error rate of an optimal stabilizer code decoder
can be effectively approximated by considering only a limited subset
$\mathcal{E}$ of the $4^{n}$ possible Pauli errors, and we have
outlined how to construct $\mathcal{E}$ without having to enumerate
all of these errors. Utilizing this approximate calculation, we have
demonstrated that there are a number of $[[5\leq n\leq12,1\leq k\leq3]]$
cyclic stabilizer codes that perform very well on both the biased
$XZ$ and AD channels across a range of error probabilities and biases.
We have also shown that an indication of the performance of a stabilizer
code can be obtained by considering the error rate of an associated
$[2n,n+k]$ classical code. We have used this as the basis for a hill
climbing algorithm, which we have shown to be effective at optimizing
codes for both of the asymmetric channels considered. Furthermore,
we have demonstrated that by modifying the mutation operation of this
hill climbing algorithm, it is possible to search for highly performant
codes that satisfy structure constraints. In particular, we have successfully
performed searches for codes with weight-four generators, CSS(Y) codes,
and linear codes. 

\bibliographystyle{apsrev4-1}
\bibliography{tailoredPaper}

\begingroup 
\squeezetable
\begin{table}
\caption{\label{tab:best cyc cocdes}Generators and distances for the best
performing inequivalent cyclic codes on the biased $XZ$ and AD channels.
Note that each stabilizer can be expressed using a single generator;
that is, each generator given corresponds to a different code. The
generators of codes performing well on both channel types are given
in bold. }
\begin{tabular*}{8.6cm}{@{\extracolsep{\fill}}>{\centering}m{1cm}>{\centering}m{3.5cm}>{\centering}m{3.5cm}}
\toprule 
$[[n,k]]$ & Biased $XZ$ & AD\tabularnewline
\midrule
$[[5,1]]$ & \textbf{\textit{YZIZY}}\textit{, $d=3$} & \textbf{\textit{YZIZY}}\textit{, $d=3$}\tabularnewline
\midrule
$[[6,1]]$ & \textit{YIZZIY, $d=2$} & \textit{XZZZZX, $d=2$}\\
\textit{YZZZZY, $d=2$}\tabularnewline
\midrule
$[[6,2]]$ & \textit{YZIZYI, $d=2$} & \textit{XIZIXY, $d=2$}\\
\textit{YIZIYX, $d=2$}\\
\textit{XZIZXY, $d=2$}\\
\textit{YZIZYX, $d=2$}\tabularnewline
\midrule
$[[6,3]]$ & \textbf{\textit{XZXXZX}}\textit{, $d=2$}\textbf{\textit{}}\\
\textbf{\textit{ XZZXZZ}}\textit{, $d=2$}\\
\textit{XIYXIY, $d=2$}\\
\textit{ YZIYZI, $d=2$} & \textbf{\textit{XZXXZX}}\textit{, $d=2$}\textbf{\textit{}}\\
\textbf{\textit{XZZXZZ}}\textit{, $d=2$}\\
\textit{YZYYZY, $d=2$}\\
\textit{YZZYZZ, $d=2$}\tabularnewline
\midrule
$[[7,1]]$ & \textbf{\textit{XZIZXII}}\textit{, $d=3$} & \textbf{\textit{XZIZXII}}\textit{, $d=3$}\\
\textit{YZIZYII , $d=3$}\tabularnewline
\midrule
$[[7,3]]$ & \textbf{\textit{XZZZXZX}}\textit{, $d=2$}\\
\textit{XZIIYZY, $d=2$}\\
\textit{YIIZYZX, $d=2$} & \textbf{\textit{XZZZXZX}}\textit{, $d=2$}\\
\textit{YZZZYZY, $d=2$}\tabularnewline
\midrule
$[[8,1]]$ & \textbf{\textit{YIIZIZZX}}\textit{, $d=3$}\\
\textit{ZZYIIIIY, $d=3$} & \textbf{\textit{YIIZIZZX}}\textit{, $d=3$}\\
\textit{XIIZIZZY, $d=3$}\tabularnewline
\midrule
$[[8,2]]$ & \textbf{\textit{YIIXIIYX}}\textit{, $d=2$}\textbf{\textit{}}\\
\textbf{\textit{YIZZIIXZ}}\textit{, $d=2$}\textbf{\textit{}}\\
\textbf{\textit{XIIYZIYY}}\textit{, $d=2$}\\
\textit{YIIZIIYZ, $d=2$}\\
\textit{YZIZIZYZ, $d=2$}\\
\textit{XZZZZZXZ, $d=2$} & \textbf{\textit{YIIXIIYX}}\textit{, $d=2$}\textbf{\textit{}}\\
\textbf{\textit{YIZZIIXZ}}\textit{, $d=2$}\textbf{\textit{}}\\
\textbf{\textit{XIIYZIYY}}\textit{, $d=2$}\\
\textit{XIIYIIXY, $d=2$}\\
\textit{YIIXZIXX, $d=2$}\\
\textit{XIZZIIYZ, $d=2$}\tabularnewline
\midrule
$[[8,3]]$ & \textbf{\textit{YIXIIYZY}}\textit{, $d=2$}\textbf{\textit{}}\\
\textbf{\textit{XZIIZXYY}}\textit{, $d=2$}\\
\textit{YZIZIXYX, $d=2$} & \textbf{\textit{YIXIIYZY}}\textit{, $d=2$}\textbf{\textit{}}\\
\textbf{\textit{XZIIZXYY}}\textit{, $d=2$}\\
\textit{XIYIIXZX, $d=2$}\\
\textit{YZIIZYXX, $d=2$}\tabularnewline
\midrule
$[[9,1]]$ & \textbf{\textit{ZIZYIIIIY}}\textit{, $d=3$} & \textbf{\textit{ZIZYIIIIY}}\textit{, $d=3$}\\
\textit{ZIZXIIIIX, $d=3$}\tabularnewline
\midrule
$[[9,2]]$ & \textbf{\textit{IZIXIZIYY}}\textit{, $d=3$} & \textbf{\textit{IZIXIZIYY}}\textit{, $d=3$}\\
\textit{IZIYIZIXX, $d=3$}\tabularnewline
\midrule
$[[9,3]]$ & \textbf{\textit{YZZIZZYII}}\textit{, $d=3$} & \textbf{\textit{YZZIZZYII}}\textit{, $d=3$}\\
\textit{XZZIZZXII , $d=3$}\tabularnewline
\midrule
$[[10,1]]$ & \textbf{\textit{YZIZIIZIZY}}\textit{, $d=4$} & \textbf{\textit{YZIZIIZIZY}}\textit{, $d=4$}\\
\textit{XZIZIIZIZX, $d=4$}\tabularnewline
\midrule
$[[10,2]]$ & \textit{YZZIIIZZYI, $d=2$} & \textit{IYXIIIIIXY, $d=3$}\tabularnewline
\midrule
$[[11,1]]$ & \textbf{\textit{IYIIZIIZIIY}}\textit{, $d=3$} & \textbf{\textit{IYIIZIIZIIY}}\textit{, $d=3$}\\
\textit{IXIIZIIZIIX, $d=3$}\tabularnewline
\midrule
$[[12,1]]$ & \textbf{\textit{YIXIXIIIIIZX}}\textit{, $d=4$} & \textbf{\textit{YIXIXIIIIIZX}}\textit{, $d=4$}\\
\textit{XIYIYIIIIIZY, $d=4$}\tabularnewline
\midrule
$[[12,2]]$ & \textbf{\textit{IIZZIIXZZIXY}}\textit{, $d=4$}\\
\textit{YXZIXIIIIIYX, $d=4$} & \textbf{\textit{IIZZIIXZZIXY}}\textit{, $d=4$}\\
\textit{IIZZIIYZZIYX, $d=4$}\tabularnewline
\midrule
$[[12,3]]$ & \textbf{\textit{ZZXIYIIIIYIX}}\textit{, $d=3$}\\
\textit{IZZIXIZZIYXY, $d=3$} & \textbf{\textit{ZZXIYIIIIYIX}}\textit{, $d=3$}\\
\textit{ZZYIXIIIIXIY, $d=3$}\tabularnewline
\bottomrule
\end{tabular*}
\end{table}

\begin{table}
\caption{\label{tab:XZ HC codes}Generators and distances for the best codes
found for the biased $XZ$ channel using hill climbing}

\begin{tabular*}{8.6cm}{@{\extracolsep{\fill}}>{\centering}m{0.5cm}>{\centering}m{2.5cm}>{\centering}m{2.5cm}>{\centering}m{2.5cm}}
\toprule
$n \backslash k$ & $1$ & $2$ & $3$
\tabularnewline
\midrule
$5$ &
\textit{IXXZZ} \\ 
\textit{YZYIZ} \\ 
\textit{IZZYY} \\ 
\textit{XZIZX}
\\ $d=3$  
&
\textit{XXYZI} \\ 
\textit{XXZXZ} \\ 
\textit{XIIZY}
\\ $d=1$  
&
\textit{XYZYZ} \\ 
\textit{IIIXX}
\\ $d=1$  
\tabularnewline
\midrule
$6$ &
\textit{IXXXYZ} \\ 
\textit{YIYIZZ} \\ 
\textit{IYYIII} \\ 
\textit{XZXIYX} \\ 
\textit{ZZXYIZ}
\\ $d=3$  
&
\textit{ZIZYXY} \\ 
\textit{XXZXYX} \\ 
\textit{IYYZYZ} \\ 
\textit{XZIZYX}
\\ $d=2$  
&
\textit{IIYYZY} \\ 
\textit{XXZYZZ} \\ 
\textit{XIIIYX}
\\ $d=1$  
\tabularnewline
\midrule
$7$ &
\textit{XZZXYYI} \\ 
\textit{XYYXZXY} \\ 
\textit{XIXZZZZ} \\ 
\textit{YZIZXXY} \\ 
\textit{ZYIXYIY} \\ 
\textit{YIYZYZY}
\\ $d=3$  
&
\textit{IIZYXZY} \\ 
\textit{ZIXYIXY} \\ 
\textit{XIZZYXZ} \\ 
\textit{XYIXYIY} \\ 
\textit{ZXZXYII}
\\ $d=2$  
&
\textit{YZZYIXX} \\ 
\textit{ZYXZIZY} \\ 
\textit{XYIXXIY} \\ 
\textit{ZZXXZXZ}
\\ $d=2$  
\tabularnewline
\midrule
$8$ &
\textit{ZIIYXIYX} \\ 
\textit{ZYYIYYXX} \\ 
\textit{IZYIYXXI} \\ 
\textit{XYXZXZII} \\ 
\textit{YIYYZIZY} \\ 
\textit{ZXXXYIII} \\ 
\textit{YXXZYYZX}
\\ $d=3$  
&
\textit{YYIIIXYI} \\ 
\textit{ZIXYZZIY} \\ 
\textit{ZYXZYXXX} \\ 
\textit{IZXYIZYI} \\ 
\textit{YZZZIZIX} \\ 
\textit{IXXIZIXI}
\\ $d=2$  
&
\textit{XZZZYYIY} \\ 
\textit{YZIYXIXX} \\ 
\textit{XZXXYYYI} \\ 
\textit{IYZYIYZI} \\ 
\textit{YYZZZXIY}
\\ $d=2$  
\tabularnewline
\midrule
$9$ &
\textit{YXYXIIIII} \\ 
\textit{YZIYIZIZI} \\ 
\textit{XYIIYIIIX} \\ 
\textit{IXIXIZZZZ} \\ 
\textit{XYXYIXZXI} \\ 
\textit{XIXXXYYYY} \\ 
\textit{XXYYYYXXZ} \\ 
\textit{YZXXZYZIZ}
\\ $d=3$  
&
\textit{XIZZZZXYX} \\ 
\textit{IIZIYYIIX} \\ 
\textit{IXYYZZYYZ} \\ 
\textit{IIZXXXXZI} \\ 
\textit{ZZXIXXIXY} \\ 
\textit{YXXZIZIYI} \\ 
\textit{ZYYYIXZXI}
\\ $d=2$  
&
\textit{ZIZZXXYIZ} \\ 
\textit{ZIIYXIIYY} \\ 
\textit{IYZXXXIIX} \\ 
\textit{ZXZYYZZZI} \\ 
\textit{YZYYIYYXI} \\ 
\textit{YYZIZZXII}
\\ $d=2$  
\tabularnewline
\midrule
$10$ &
\textit{XYYXYIXYXX} \\ 
\textit{XYYYIZXZYY} \\ 
\textit{ZIXZXYIZYI} \\ 
\textit{YXYYXXIIIX} \\ 
\textit{IZZZIIZYYZ} \\ 
\textit{XXZXYZYXYZ} \\ 
\textit{XXIZXIXXIX} \\ 
\textit{ZXXIYYIXIX} \\ 
\textit{XZXZYYXXXI}
\\ $d=3$  
&
\textit{ZXZIXZIIIY} \\ 
\textit{YIIXYZIZII} \\ 
\textit{XIYIIIZXZY} \\ 
\textit{ZIXIYYIIYX} \\ 
\textit{XZZYIXIYXX} \\ 
\textit{YYIXYXYXII} \\ 
\textit{ZZYZZXXZXI} \\ 
\textit{ZXZYXZYYZI}
\\ $d=3$  
&
\textit{ZZXXIIXZZY} \\ 
\textit{IYZYZZYIZY} \\ 
\textit{YZIYXZZZIX} \\ 
\textit{ZXYZIIYXXY} \\ 
\textit{ZYXIXZIYZY} \\ 
\textit{XZYXYYIYXY} \\ 
\textit{YYYYIZXIII}
\\ $d=2$  
\tabularnewline
\midrule
$11$ &
\textit{IZXZXZXXIZX} \\ 
\textit{ZXIZXYIIIYY} \\ 
\textit{ZXYIIXYYYXI} \\ 
\textit{YIYZXXZIYXX} \\ 
\textit{IYZYXXIYZYX} \\ 
\textit{IYXXYXIYXZZ} \\ 
\textit{ZIXZYIZXZIX} \\ 
\textit{YYXZXYXYZXX} \\ 
\textit{ZIYZXYXZIYI} \\ 
\textit{YYZIXZZIZIZ}
\\ $d=3$  
&
\textit{YZXIXIZYXZX} \\ 
\textit{ZXIIYIXYZXI} \\ 
\textit{YYXZXYIIXXX} \\ 
\textit{ZXIIZXZYYZX} \\ 
\textit{IXZYIIXYIIZ} \\ 
\textit{ZXIZIXYXIXY} \\ 
\textit{IZIXZZZYXXX} \\ 
\textit{ZYXIZYXXXXX} \\ 
\textit{XZIZIIIXYIY}
\\ $d=3$  
&
\textit{YZXZXXZXXYY} \\ 
\textit{YZIYXIIYIXX} \\ 
\textit{YYIZYXYZXXX} \\ 
\textit{ZYIZIIYYZYX} \\ 
\textit{YYZIIYXIZXY} \\ 
\textit{YXZXYXZZYII} \\ 
\textit{YIIIIZYZIYX} \\ 
\textit{IZXXYXZIYIX}
\\ $d=2$  
\tabularnewline
\midrule
$12$ &
\textit{YYIIXIYYIXZZ} \\ 
\textit{YXIYZIXZIZIY} \\ 
\textit{XXXXIZXIZXXZ} \\ 
\textit{ZXZYIIIIZYXI} \\ 
\textit{ZYYYZIZIYIZX} \\ 
\textit{IZYYXXXYIYIY} \\ 
\textit{ZYYYYIYYXIYX} \\ 
\textit{ZYIZYXIXIYXY} \\ 
\textit{IZIXIXYXXYYZ} \\ 
\textit{IXXZZYIIXXXY} \\ 
\textit{ZZIZYXIZXYZX}
\\ $d=4$  
&
\textit{ZXZXZYXXZZYI} \\ 
\textit{ZIZYZIXIXZIY} \\ 
\textit{IYYYXZXXZYYY} \\ 
\textit{IXIIIYYZIZYZ} \\ 
\textit{IIYIIXYYZIIZ} \\ 
\textit{IIIXYYIZXYIX} \\ 
\textit{ZIXYIIZXIZZY} \\ 
\textit{YIYZZXYZIXXX} \\ 
\textit{YYXXYXXZYYYI} \\ 
\textit{YZYXZZYYYYII}
\\ $d=3$  
&
\textit{IIYYYIZXIXII} \\ 
\textit{IZXZZXXIIXZZ} \\ 
\textit{ZZZYZXYIZZII} \\ 
\textit{YZIXIZXIIIXX} \\ 
\textit{XYZZZIXZYYYI} \\ 
\textit{XIYZYXIXXYIY} \\ 
\textit{YIZZYXXYZIII} \\ 
\textit{ZYXZXIIZIIZI} \\ 
\textit{ZXZZYYXZXIYY}
\\ $d=3$  
\tabularnewline
\bottomrule
\end{tabular*}
\end{table}

\begin{table}
\caption{\label{tab:AD HC codes}Generators and distances for the best codes
found for the AD channel using hill climbing.}

\begin{tabular*}{8.6cm}{@{\extracolsep{\fill}}>{\centering}m{0.5cm}>{\centering}m{2.5cm}>{\centering}m{2.5cm}>{\centering}m{2.5cm}}
\toprule
$n \backslash k$ & $1$ & $2$ & $3$
\tabularnewline
\midrule
$5$ &
\textit{YYXIX} \\ 
\textit{IXYYX} \\ 
\textit{IZXXZ} \\ 
\textit{XZIZX}
\\ $d=3$  
&
\textit{YXZZX} \\ 
\textit{IIYYX} \\ 
\textit{IYIYX}
\\ $d=1$  
&
\textit{XZZYX} \\ 
\textit{IXIXI}
\\ $d=1$  
\tabularnewline
\midrule
$6$ &
\textit{XZZZIY} \\ 
\textit{IZYXYY} \\ 
\textit{YIZXZI} \\ 
\textit{YYXIXZ} \\ 
\textit{IZZYXZ}
\\ $d=3$  
&
\textit{XYIYZY} \\ 
\textit{XYYIXZ} \\ 
\textit{IIXXZZ} \\ 
\textit{XZYZZI}
\\ $d=1$  
&
\textit{YIXIYZ} \\ 
\textit{IYXYYI} \\ 
\textit{IZXYZY}
\\ $d=1$  
\tabularnewline
\midrule
$7$ &
\textit{XZZIYIX} \\ 
\textit{XIYZYXZ} \\ 
\textit{IYYXXXZ} \\ 
\textit{ZIXXIYI} \\ 
\textit{XZZZIZI} \\ 
\textit{XIXYZIX}
\\ $d=3$  
&
\textit{ZYYXYXX} \\ 
\textit{XYZZIZI} \\ 
\textit{XZIZZYZ} \\ 
\textit{YZZXIXY} \\ 
\textit{XYIYXXZ}
\\ $d=2$  
&
\textit{IZXIZYX} \\ 
\textit{YZYXIIZ} \\ 
\textit{XYIIZIZ} \\ 
\textit{IYYXXIY}
\\ $d=1$  
\tabularnewline
\midrule
$8$ &
\textit{ZXXIYIYX} \\ 
\textit{YXXZXIIX} \\ 
\textit{YIYZIIXZ} \\ 
\textit{IXXXXYIZ} \\ 
\textit{ZYXZIZZX} \\ 
\textit{YIZXYZZY} \\ 
\textit{IYXZYZZI}
\\ $d=3$  
&
\textit{YIYYYXIZ} \\ 
\textit{YYXXZYXZ} \\ 
\textit{ZYXIXYIZ} \\ 
\textit{ZZZXXIIZ} \\ 
\textit{YZIIXXXY} \\ 
\textit{IXIIXIZY}
\\ $d=3$  
&
\textit{XZXIZYZI} \\ 
\textit{ZXXYXXZY} \\ 
\textit{YZZIXXIZ} \\ 
\textit{XIXXXZIX} \\ 
\textit{IZXYIZYZ}
\\ $d=3$  
\tabularnewline
\midrule
$9$ &
\textit{ZIXYXIYYI} \\ 
\textit{ZYYZXXYIX} \\ 
\textit{IYIXIYIZI} \\ 
\textit{YYZZIXIZI} \\ 
\textit{IXYYYXYYY} \\ 
\textit{IIIZIXZIY} \\ 
\textit{ZXZYIYYZX} \\ 
\textit{ZZXZZXIYY}
\\ $d=3$  
&
\textit{IIYYXYIYI} \\ 
\textit{XZYXXZXII} \\ 
\textit{IXIZXXYIZ} \\ 
\textit{YZXXIYZYI} \\ 
\textit{YXZYXYZXY} \\ 
\textit{YZYYZIYZX} \\ 
\textit{YYYIYZYIX}
\\ $d=3$  
&
\textit{IXIYZYZZI} \\ 
\textit{YZIIYZZXZ} \\ 
\textit{ZYIZIYIIX} \\ 
\textit{YZXXIXYII} \\ 
\textit{YXIIIIXYX} \\ 
\textit{YZZIXYIZX}
\\ $d=3$  
\tabularnewline
\midrule
$10$ &
\textit{YIIZYYXYYY} \\ 
\textit{XYZXIZZXYZ} \\ 
\textit{IXYXZYYYYX} \\ 
\textit{XYIIZZXXXZ} \\ 
\textit{XYIYIYXXIX} \\ 
\textit{XZZYZYYIXX} \\ 
\textit{ZIXIYZYZII} \\ 
\textit{IIXXXIZIXI} \\ 
\textit{XXYYXXXXZI}
\\ $d=3$  
&
\textit{YZXYZYYIYY} \\ 
\textit{YYXZIIXZXI} \\ 
\textit{XIIYIIXXYI} \\ 
\textit{IYZZIIXXXX} \\ 
\textit{YXZIIXZIYY} \\ 
\textit{XXYYZIZZYI} \\ 
\textit{YYZYZZYIYZ} \\ 
\textit{YXYXXIZYYX}
\\ $d=3$  
&
\textit{ZZZZYZZXZX} \\ 
\textit{XXZXZZXXYX} \\ 
\textit{XXYZXZXZXY} \\ 
\textit{YIZYIXZXYX} \\ 
\textit{ZXYYZXIYYY} \\ 
\textit{XIXYXZIZYX} \\ 
\textit{IIIZYXIYXZ}
\\ $d=3$  
\tabularnewline
\midrule
$11$ &
\textit{ZXYZXYIZZXX} \\ 
\textit{XZXZYIXXYZY} \\ 
\textit{YZXZIZYXXYY} \\ 
\textit{YYYIXZIIYZY} \\ 
\textit{ZZZZXZIXZYZ} \\ 
\textit{IZZXIIIXZXI} \\ 
\textit{ZZIXYZXZYXY} \\ 
\textit{IZZIYXIZXXY} \\ 
\textit{XYZZXZZXZIZ} \\ 
\textit{YIYYIIYIZXI}
\\ $d=3$  
&
\textit{XIYZIZIIXZZ} \\ 
\textit{YZZYXIYXYYY} \\ 
\textit{IYIIZIYZYZY} \\ 
\textit{ZZIYZZYZIIZ} \\ 
\textit{IZYYZXXYYZX} \\ 
\textit{YYZZIYYXZYX} \\ 
\textit{ZZZZYZYXYIX} \\ 
\textit{ZXYIXXZYXYY} \\ 
\textit{IIZXIYYYYYY}
\\ $d=3$  
&
\textit{ZIYIZXXXIZZ} \\ 
\textit{XZIYXZIIXZX} \\ 
\textit{IYZZIXIYXYX} \\ 
\textit{ZYIYIYZZXXI} \\ 
\textit{YXXZYIIXZYI} \\ 
\textit{YYZYYZXYYZX} \\ 
\textit{IYYXXXIZIXX} \\ 
\textit{YXXXYXIXIZY}
\\ $d=3$  
\tabularnewline
\midrule
$12$ &
\textit{IXIIZZIZIXYI} \\ 
\textit{XXIZXXIIIXII} \\ 
\textit{ZXXYIZYXIZYY} \\ 
\textit{ZYYYYIXZYXIX} \\ 
\textit{YZYYXYXZZXZI} \\ 
\textit{IZIZZYIXIZIZ} \\ 
\textit{ZZIXYYIZYYYZ} \\ 
\textit{YYIXZXZYIIZZ} \\ 
\textit{ZZXXXXZZXYYY} \\ 
\textit{ZZIIXXZIZIXI} \\ 
\textit{XYXZIYZXIXZZ}
\\ $d=3$  
&
\textit{IIYXXIXZYIYI} \\ 
\textit{ZIXXZIZYIIIX} \\ 
\textit{YZZZZIYYIIZI} \\ 
\textit{IZXIYYZZXYII} \\ 
\textit{ZZYYIYZXYZXZ} \\ 
\textit{IYXIXYYYIYXX} \\ 
\textit{XYXYZXYIXYZZ} \\ 
\textit{XIXIYIZZZZYY} \\ 
\textit{XXYXXXXYXIIX} \\ 
\textit{YXZYXIIIZIXI}
\\ $d=3$  
&
\textit{XYZYZIZIIXXX} \\ 
\textit{XXYIYIIZIIXZ} \\ 
\textit{YZZZYYYXIZII} \\ 
\textit{ZIIIXYXYXIYI} \\ 
\textit{XIZZZYYIXIZY} \\ 
\textit{IXXIIZXYXYXI} \\ 
\textit{ZZXIIYZZYIZI} \\ 
\textit{ZYZZIZXIIZZZ} \\ 
\textit{ZZXIYYZXIYXZ}
\\ $d=3$  
\tabularnewline
\bottomrule
\end{tabular*}
\end{table}

\begin{table}
\caption{\label{tab:XZ HC codes w4}Generators and distances for the best weight-four
codes found for the biased $XZ$ channel using hill climbing.}

\begin{tabular*}{8.6cm}{@{\extracolsep{\fill}}>{\centering}m{0.5cm}>{\centering}m{2.5cm}>{\centering}m{2.5cm}>{\centering}m{2.5cm}}
\toprule
$n \backslash k$ & $1$ & $2$ & $3$
\tabularnewline
\midrule
$5$ &
\textit{ZIXZX} \\ 
\textit{IZXXZ} \\ 
\textit{ZYYIZ} \\ 
\textit{XZZIX}
\\ $d=3$  
&
\textit{XYYIX} \\ 
\textit{ZXIXX} \\ 
\textit{ZYYIZ}
\\ $d=1$  
&
\textit{XIYZZ} \\ 
\textit{IXYZZ}
\\ $d=1$  
\tabularnewline
\midrule
$6$ &
\textit{IIXZZX} \\ 
\textit{YIIYXX} \\ 
\textit{ZZIIXY} \\ 
\textit{IXZIXX} \\ 
\textit{YIZIZY}
\\ $d=2$  
&
\textit{XIZXYI} \\ 
\textit{IYYXXI} \\ 
\textit{IZIZXX} \\ 
\textit{IZYIZY}
\\ $d=1$  
&
\textit{IXYYIX} \\ 
\textit{IIXZYX} \\ 
\textit{XXYIXI}
\\ $d=1$  
\tabularnewline
\midrule
$7$ &
\textit{IZXIIZY} \\ 
\textit{ZZIXXII} \\ 
\textit{YIXZIZI} \\ 
\textit{IXIZXIZ} \\ 
\textit{IZZIIXY} \\ 
\textit{ZYYIZII}
\\ $d=3$  
&
\textit{IIYIYZX} \\ 
\textit{IZIXIYY} \\ 
\textit{ZYIIYIX} \\ 
\textit{XIIYXXI} \\ 
\textit{YIZXIIY}
\\ $d=2$  
&
\textit{YIZIIYY} \\ 
\textit{XIXYIIY} \\ 
\textit{IZYIXIZ} \\ 
\textit{IYIZIZX}
\\ $d=1$  
\tabularnewline
\midrule
$8$ &
\textit{IZZIYIYI} \\ 
\textit{IIIYYIYY} \\ 
\textit{XXYIIIIY} \\ 
\textit{YIIXIXIX} \\ 
\textit{IIYZZYII} \\ 
\textit{YYIXXIII} \\ 
\textit{IIIIXZZY}
\\ $d=3$  
&
\textit{XIIZXIIX} \\ 
\textit{YIIYIXXI} \\ 
\textit{IIYXZYII} \\ 
\textit{XIIIZIYY} \\ 
\textit{IYXIIZYI} \\ 
\textit{ZXIIIYIZ}
\\ $d=2$  
&
\textit{IYIYIYYI} \\ 
\textit{IIIYXIYY} \\ 
\textit{XIIIYXXI} \\ 
\textit{YYXIIIYI} \\ 
\textit{IXYXIIIZ}
\\ $d=2$  
\tabularnewline
\midrule
$9$ &
\textit{IXYYIXIII} \\ 
\textit{ZIIYZIIIY} \\ 
\textit{IIIIYIZYZ} \\ 
\textit{ZIYIIIYZI} \\ 
\textit{IIZYIZIYI} \\ 
\textit{IZIIIYIZY} \\ 
\textit{YIYIIIIXX} \\ 
\textit{IYIXYXIII}
\\ $d=3$  
&
\textit{YIXIIIXIX} \\ 
\textit{YYIIZXIII} \\ 
\textit{XIIIXIYYI} \\ 
\textit{IIXXIIXXI} \\ 
\textit{XIZIIYIYI} \\ 
\textit{IIYIZIIZY} \\ 
\textit{IXIYXIIYI}
\\ $d=2$  
&
\textit{XXXIXIIII} \\ 
\textit{YYIIIZIXI} \\ 
\textit{ZIYIIYXII} \\ 
\textit{YIIYZIYII} \\ 
\textit{IIYZYIIIY} \\ 
\textit{IIIYIYIYZ}
\\ $d=2$  
\tabularnewline
\midrule
$10$ &
\textit{XIIXYIIXII} \\ 
\textit{IIYYZIXIII} \\ 
\textit{IIIXIIYIZY} \\ 
\textit{IIZIZIYYII} \\ 
\textit{ZIIIIXIYIY} \\ 
\textit{IZYIIZIXII} \\ 
\textit{IYIXIXYIII} \\ 
\textit{IZIIYIZIYI} \\ 
\textit{YZIYIIIIIZ}
\\ $d=3$  
&
\textit{IIIIZYIXIY} \\ 
\textit{XIIIIIYIXY} \\ 
\textit{IIIYZIZIYI} \\ 
\textit{IYXIIIXIZI} \\ 
\textit{XIYXIIYIII} \\ 
\textit{YIXIIXIIIX} \\ 
\textit{IXIIYXYIII} \\ 
\textit{YIIIXIXYII}
\\ $d=2$  
&
\textit{XIIIIYIYIX} \\ 
\textit{IXYXIIYIII} \\ 
\textit{XYXIYIIIII} \\ 
\textit{XYIYIIIIYI} \\ 
\textit{IIIYIXZIIY} \\ 
\textit{IIYIXYYIII} \\ 
\textit{ZIYIIIIXXI}
\\ $d=2$  
\tabularnewline
\midrule
$11$ &
\textit{IIIIYZIZYII} \\ 
\textit{IZYYYIIIIII} \\ 
\textit{YIIIYIIXXII} \\ 
\textit{IIIYIIZIIZY} \\ 
\textit{IIYIIIZIIYZ} \\ 
\textit{XIIIIIYIYXI} \\ 
\textit{YYIZIIIIIYI} \\ 
\textit{ZIIYIYIIZII} \\ 
\textit{IIXIXIIYIIY} \\ 
\textit{IYXIIIXIIYI}
\\ $d=3$  
&
\textit{IIIIIYZYIZI} \\ 
\textit{IZIYIIIYIIX} \\ 
\textit{IIIIYXIIZYI} \\ 
\textit{YXIIIIXXIII} \\ 
\textit{XIXIIIYIIYI} \\ 
\textit{XYIIIIIIZIY} \\ 
\textit{YXYXIIIIIII} \\ 
\textit{IXIXXIIIYII} \\ 
\textit{IIIIIIXIYXX}
\\ $d=2$  
&
\textit{IIYYIYIIZII} \\ 
\textit{YIIIXIYIIXI} \\ 
\textit{IYXZIIIIIIY} \\ 
\textit{XIIZYIIIYII} \\ 
\textit{YZIYIIIYIII} \\ 
\textit{XIIIIIIXIYY} \\ 
\textit{IIXIYXXIIII} \\ 
\textit{YIYIXIIIIIX}
\\ $d=2$  
\tabularnewline
\midrule
$12$ &
\textit{IZIIIIYZYIII} \\ 
\textit{IIIIZYYIIZII} \\ 
\textit{IYXIIIIIZIIY} \\ 
\textit{XIIIIYIIYIIX} \\ 
\textit{IYIZXIZIIIII} \\ 
\textit{YXIXIIIIXIII} \\ 
\textit{IXZIIIXIIYII} \\ 
\textit{IIZIIIIYXIIY} \\ 
\textit{XIIYZIIIIIXI} \\ 
\textit{YIIIIZIIIYZI} \\ 
\textit{ZIIIIIIIYIYZ}
\\ $d=3$  
&
\textit{IIYXIIIYIIIZ} \\ 
\textit{IIXIZYIIIIIY} \\ 
\textit{ZIIYIIIIIYIY} \\ 
\textit{IIIIIIYIYYZI} \\ 
\textit{YIIZIIIZYIII} \\ 
\textit{IIIIYZXIIIYI} \\ 
\textit{YIIIIYZIIXII} \\ 
\textit{IIIIIIIXXIYY} \\ 
\textit{IYIIXXIIIYII} \\ 
\textit{IXYIXYIIIIII}
\\ $d=2$  
&
\textit{IZZIIIYIIIIY} \\ 
\textit{XIIXIZYIIIII} \\ 
\textit{YIIYYIIYIIII} \\ 
\textit{IIIYIYIIYIZI} \\ 
\textit{IYIIIIIYYIIX} \\ 
\textit{IIYIIXXIIIYI} \\ 
\textit{IIIIXIIXZIYI} \\ 
\textit{YIZIIYIIIYII} \\ 
\textit{IXYIIIIIIZIY}
\\ $d=2$  
\tabularnewline
\bottomrule
\end{tabular*}
\end{table}

\begin{table}
\caption{\label{tab:AD HC codes w4}Generators and distances for the best weight-four
codes found for the AD channel using hill climbing.}

\begin{tabular*}{8.6cm}{@{\extracolsep{\fill}}>{\centering}m{0.5cm}>{\centering}m{2.5cm}>{\centering}m{2.5cm}>{\centering}m{2.5cm}}
\toprule
$n \backslash k$ & $1$ & $2$ & $3$
\tabularnewline
\midrule
$5$ &
\textit{IXZXZ} \\ 
\textit{YZIYZ} \\ 
\textit{IZYZY} \\ 
\textit{ZXIZX}
\\ $d=3$  
&
\textit{ZXYIY} \\ 
\textit{ZXYYI} \\ 
\textit{XIYXX}
\\ $d=1$  
&
\textit{XYIXZ} \\ 
\textit{XYYIZ}
\\ $d=1$  
\tabularnewline
\midrule
$6$ &
\textit{YIXXYI} \\ 
\textit{YXIIYZ} \\ 
\textit{ZIXIXZ} \\ 
\textit{IXIYZY} \\ 
\textit{XIZZXI}
\\ $d=1$  
&
\textit{YIYXXI} \\ 
\textit{IZYIXY} \\ 
\textit{YIZIYY} \\ 
\textit{ZXIYIZ}
\\ $d=2$  
&
\textit{IYYXYI} \\ 
\textit{XXYZII} \\ 
\textit{ZYIIYX}
\\ $d=1$  
\tabularnewline
\midrule
$7$ &
\textit{ZYIIXXI} \\ 
\textit{ZIIYIXZ} \\ 
\textit{XIIXYIZ} \\ 
\textit{YZYZIII} \\ 
\textit{IYXZIIY} \\ 
\textit{IIZXYYI}
\\ $d=3$  
&
\textit{ZYIIXIX} \\ 
\textit{IXYIYYI} \\ 
\textit{ZIXZXII} \\ 
\textit{YXIYIYI} \\ 
\textit{IIIXZXY}
\\ $d=2$  
&
\textit{XYYIXII} \\ 
\textit{XIIXXXI} \\ 
\textit{ZIXYIIY} \\ 
\textit{ZZIIXYI}
\\ $d=1$  
\tabularnewline
\midrule
$8$ &
\textit{ZIYIIYXI} \\ 
\textit{IZXYIZII} \\ 
\textit{YIIIZIZX} \\ 
\textit{IIIZIXYY} \\ 
\textit{YYIIIYZI} \\ 
\textit{IIXIXIZZ} \\ 
\textit{IIYYZYII}
\\ $d=3$  
&
\textit{XIYIIYYI} \\ 
\textit{IYIIIZXX} \\ 
\textit{IIXIXXIY} \\ 
\textit{IYYXYIII} \\ 
\textit{XXIZIIIY} \\ 
\textit{YIIYXIXI}
\\ $d=3$  
&
\textit{IXYXIIIZ} \\ 
\textit{YIIXYYII} \\ 
\textit{IIIZIXYX} \\ 
\textit{IYIIZZIY} \\ 
\textit{IZIIYIZY}
\\ $d=1$  
\tabularnewline
\midrule
$9$ &
\textit{IIYIIXYYI} \\ 
\textit{IIIXXXIIX} \\ 
\textit{IYXYIYIII} \\ 
\textit{XYXIIIIXI} \\ 
\textit{IIXYIIXIY} \\ 
\textit{YIYIIXIIX} \\ 
\textit{IXYIXIYII} \\ 
\textit{XIIIYYXII}
\\ $d=3$  
&
\textit{XIXIIXIYI} \\ 
\textit{IIIYIZXXI} \\ 
\textit{IIYIXIIXY} \\ 
\textit{YIIXYIIZI} \\ 
\textit{YYIIIYYII} \\ 
\textit{IXIXIIZIY} \\ 
\textit{IXXIIIXIX}
\\ $d=2$  
&
\textit{YIYIIIYIZ} \\ 
\textit{XXIIIXIIX} \\ 
\textit{IXXIZIXII} \\ 
\textit{XIZIXIIXI} \\ 
\textit{YIIXIYIZI} \\ 
\textit{IYIIXIXIY}
\\ $d=1$  
\tabularnewline
\midrule
$10$ &
\textit{YIIYYIIIIY} \\ 
\textit{IIIXIXIXIX} \\ 
\textit{IYYIIIIYIY} \\ 
\textit{XXXIXIIIII} \\ 
\textit{IYIIYIYIZI} \\ 
\textit{IIIXXXIIXI} \\ 
\textit{ZIYIXIXIII} \\ 
\textit{IIIIIZXIYY} \\ 
\textit{XIXXIIIXII}
\\ $d=3$  
&
\textit{IIYXIYYIII} \\ 
\textit{IIIIYXZYII} \\ 
\textit{IXZIXIXIII} \\ 
\textit{IYYIIIIIXY} \\ 
\textit{YZIIYIIIIX} \\ 
\textit{IIIYIXIIZX} \\ 
\textit{IIIZIIXXXI} \\ 
\textit{XIIIXYIIXI}
\\ $d=3$  
&
\textit{IIIIZXXIXI} \\ 
\textit{YYXIIIIIXI} \\ 
\textit{XIYZIIIIIX} \\ 
\textit{YIIXIXIYII} \\ 
\textit{IIIIYYIXIX} \\ 
\textit{IIZIXIIIYY} \\ 
\textit{IIIYIZYIIY}
\\ $d=1$  
\tabularnewline
\midrule
$11$ &
\textit{YIYIIYIIIIY} \\ 
\textit{IIXIYZIYIII} \\ 
\textit{IXIYIIZYIII} \\ 
\textit{IXYIXIIIIXI} \\ 
\textit{IIZIIXIIYYI} \\ 
\textit{IIIXIYIXIXI} \\ 
\textit{XIIXIIXIIIX} \\ 
\textit{ZXIIIIIIYIZ} \\ 
\textit{YIIIIIZIXXI} \\ 
\textit{IYZIIIIXIIX}
\\ $d=3$  
&
\textit{IIIIIIXIYXY} \\ 
\textit{IIIZIIYXIIX} \\ 
\textit{IYIYIXIZIII} \\ 
\textit{IYZIIIIIXYI} \\ 
\textit{YIXIYIIIYII} \\ 
\textit{IXIIYIIXYII} \\ 
\textit{ZIYIIXIIIXI} \\ 
\textit{IIIIXXYIXII} \\ 
\textit{YIIXIZXIIII}
\\ $d=3$  
&
\textit{YIIIIIXIXIY} \\ 
\textit{XIIIIYIXZII} \\ 
\textit{IIXYXIIIIIY} \\ 
\textit{IYIIXIIXIZI} \\ 
\textit{IIXIIZIYIXI} \\ 
\textit{IXIXZIXIIII} \\ 
\textit{IIZXIXIIXII} \\ 
\textit{XZIIIIYIIXI}
\\ $d=1$  
\tabularnewline
\midrule
$12$ &
\textit{IXIIYIIXXIII} \\ 
\textit{IIXIIIIXIYIX} \\ 
\textit{IIIIXIIIYXIY} \\ 
\textit{IIIIXYIYIIIY} \\ 
\textit{IIIYYXIIIYII} \\ 
\textit{ZXIIIIZIIIXI} \\ 
\textit{IZIIIIYIZYII} \\ 
\textit{IIYXIYIIIIIY} \\ 
\textit{YIIXIYIIIIYI} \\ 
\textit{YIIIIIXIXYII} \\ 
\textit{XIXYIIYIIIII}
\\ $d=3$  
&
\textit{IIYZZIIIIIYI} \\ 
\textit{IYIIIZIIIYYI} \\ 
\textit{IIXIYIIYIIIX} \\ 
\textit{IIIIXYIIIIZY} \\ 
\textit{YIIIIXIIIXIX} \\ 
\textit{YIYIIIYXIIII} \\ 
\textit{IXIYIIIIXIZI} \\ 
\textit{ZXIIIYXIIIII} \\ 
\textit{IYIIIIZYZIII} \\ 
\textit{IIIIIIIXXYIY}
\\ $d=3$  
&
\textit{IIIYXZIIXIII} \\ 
\textit{XIIYIIIIIIXX} \\ 
\textit{IIYZIXIIIIIY} \\ 
\textit{IIIIZIIIYYXI} \\ 
\textit{ZIIIIIYXIIIY} \\ 
\textit{YIIIIIIYIZYI} \\ 
\textit{IYIIIIIXXXII} \\ 
\textit{IXYIIIZYIIII} \\ 
\textit{IZZIIYIIYIII}
\\ $d=3$  
\tabularnewline
\bottomrule
\end{tabular*}
\end{table}

\begin{table}
\caption{\label{tab:best w4 cyc cocdes}Generators and distances for the best
performing inequivalent cyclic codes with weight-four generators on
the biased $XZ$ and AD channels. If a code requires two generators,
they are grouped in brackets; otherwise, a single generator is given
as in Table \ref{tab:best cyc cocdes}. The generators of codes performing
well on both channel types are given in bold, while the generators
for codes previously appearing in Table \ref{tab:best cyc cocdes}
are marked with an asterisk.}
\begin{tabular*}{8.6cm}{@{\extracolsep{\fill}}>{\centering}m{1cm}>{\centering}m{3.5cm}>{\centering}m{3.5cm}}
\toprule 
$[[n,k]]$ & Biased $XZ$ & AD\tabularnewline
\midrule
$[[5,1]]$ & \textbf{\textit{YZIZY{*}}}\textit{, $d=3$} & \textbf{\textit{YZIZY{*}}}\textit{, $d=3$}\tabularnewline
\midrule
$[[6,1]]$ & \textit{YIZZIY}\textbf{\textit{{*}}}\textit{, $d=2$} & \textit{YZIIZY, $d=2$}\\
\textit{XZIIZX, $d=2$}\tabularnewline
\midrule
$[[6,2]]$ & \textit{YZIZYI}\textbf{\textit{{*}}}\textit{, $d=2$} & \textit{XIZIXY}\textbf{\textit{{*}}}\textit{, $d=2$}\\
\textit{YIZIYX}\textbf{\textit{{*}}}\textit{, $d=2$}\tabularnewline
\midrule
$[[6,3]]$ & \textbf{\textit{XIYXIY{*}}}\textit{, $d=2$}\textbf{\textit{}}\\
\textbf{\textit{YZIYZI{*}}}\textit{, $d=2$} & \textbf{\textit{XIYXIY}}\textit{, $d=2$}\textbf{\textit{}}\\
\textbf{\textit{YZIYZI}}\textit{, $d=2$}\\
\textit{XZIXZI, $d=2$}\tabularnewline
\midrule
$[[7,1]]$ & \textbf{\textit{XZIZXII{*}}}\textit{, $d=3$} & \textbf{\textit{XZIZXII{*}}}\textit{, $d=3$}\\
\textit{YZIZYII}\textbf{\textit{{*}}}\textit{, $d=3$}\tabularnewline
\midrule
$[[8,1]]$ & \textbf{\textit{ZZYIIIIY{*}}}\textit{, $d=3$} & \textbf{\textit{ZZYIIIIY}}\textit{, $d=3$}\\
\textit{ZZXIIIIX , $d=3$}\tabularnewline
\midrule
$[[8,2]]$ & \textbf{\textit{YIIXIIYX{*}}}\textit{, $d=2$}\\
\textit{YIIZIIYZ}\textbf{\textit{{*}}}\textit{, $d=2$} & \textbf{\textit{YIIXIIYX{*}}}\textit{, $d=2$}\\
\textit{XIIYIIXY}\textbf{\textit{{*}}}\textit{, $d=2$}\tabularnewline
\midrule
$[[8,3]]$ & \textbf{\textit{YIIIIYYY}}\textit{, $d=1$}\\
\textbf{\textit{XIIIIXXX}}\textit{, $d=1$} & \textbf{\textit{YIIIIYYY}}\textit{, $d=1$}\\
\textbf{\textit{XIIIIXXX}}\textit{, $d=1$}\tabularnewline
\midrule
$[[9,1]]$ & \textbf{\textit{ZIZYIIIIY{*}}}\textit{, $d=3$} & \textbf{\textit{ZIZYIIIIY{*}}}\textit{, $d=3$}\\
\textit{ZIZXIIIIX}\textbf{\textit{{*}}}\textit{, $d=3$}\tabularnewline
\midrule
$[[9,3]]$ & \textbf{\textit{IIIYYIYYI}}\textit{, $d=1$}\textbf{\textit{}}\\
\textbf{\textit{IIIXXIXXI}}\textit{, $d=1$} & \textbf{\textit{IIIYYIYYI}}\textit{, $d=1$}\textbf{\textit{}}\\
\textbf{\textit{IIIXXIXXI}}\textit{, $d=1$}\tabularnewline
\midrule
$[[10,1]]$ & \textit{IIYZIIIIZY, $d=2$} & \textit{XIIZIIZIIX, $d=3$}\\
\textit{YIIZIIZIIY, $d=3$}\tabularnewline
\midrule
$[[10,2]]$ & \textbf{\textit{IYXIIIIIXY}}\textit{, $d=3$}\\
\textit{IZYIIIIIYZ, $d=3$} & \textbf{\textit{IYXIIIIIXY{*}}}\textit{, $d=3$}\tabularnewline
\midrule
$[[11,1]]$ & \textbf{\textit{IYIIZIIZIIY{*}}}\textit{, $d=3$} & \textbf{\textit{IYIIZIIZIIY{*}}}\textit{, $d=3$}\\
\textit{IXIIZIIZIIX}\textbf{\textit{{*}}}\textit{, $d=3$}\tabularnewline
\midrule
$[[12,1]]$ & \textbf{\textit{IIIIYIIZZIIY}}\textit{, $d=3$} & \textbf{\textit{IIIIYIIZZIIY}}\textit{, $d=3$}\\
\textit{IIIIXIIZZIIX, $d=3$}\tabularnewline
\midrule
$[[12,2]]$ & \textit{YIIIIZIIIIYZ, $d=2$} & \textit{XZIIIIIIIZXI, $d=3$}\\
\textit{YZIIIIIIIZYI, $d=3$}\tabularnewline
\midrule
$[[12,3]]$ & \textit{(}\textbf{\textit{XIIXIIXIIXII}}\textit{,~~~~~~~~~~}\\
\textbf{\textit{IYIIIIYIIYYI}}\textit{), $d=2$} & \textit{(}\textbf{\textit{XIIXIIXIIXII}}\textit{,~~~~~~~~~~}\\
\textbf{\textit{IYIIIIYIIYYI}}\textit{), $d=2$}\\
\textit{(YIIYIIYIIYII,~~~~~~~~~~}\\
\textit{IXIIIIXIIXXI), $d=2$}\tabularnewline
\bottomrule
\end{tabular*}
\end{table}

\begin{table}
\caption{\label{tab:XZ HC codes CSSY}Generators and distances for the best
CSSY codes found for the biased $XZ$ channel using hill climbing.}

\begin{tabular*}{8.6cm}{@{\extracolsep{\fill}}>{\centering}m{0.5cm}>{\centering}m{2.5cm}>{\centering}m{2.5cm}>{\centering}m{2.5cm}}
\toprule
$n \backslash k$ & $1$ & $2$ & $3$
\tabularnewline
\midrule
$5$ &
\textit{XXXXX} \\ 
\textit{IIIYY} \\ 
\textit{IYIYI} \\ 
\textit{IIYIY}
\\ $d=1$  
&
\textit{XXXIX} \\ 
\textit{YYIYI} \\ 
\textit{YIYYI}
\\ $d=1$  
&
\textit{XXXXI} \\ 
\textit{XXIXX}
\\ $d=1$  
\tabularnewline
\midrule
$6$ &
\textit{XXXXXX} \\ 
\textit{IIYYYY} \\ 
\textit{IYYIYY} \\ 
\textit{YIYYIY} \\ 
\textit{IYYYYI}
\\ $d=2$  
&
\textit{IXXXXX} \\ 
\textit{YIYYII} \\ 
\textit{YIYIYI} \\ 
\textit{YYIYII}
\\ $d=1$  
&
\textit{IXIXXX} \\ 
\textit{YYIIYI} \\ 
\textit{YIYIYY}
\\ $d=1$  
\tabularnewline
\midrule
$7$ &
\textit{IIIXIXX} \\ 
\textit{XXXXXIX} \\ 
\textit{YYIYYYI} \\ 
\textit{YIIYYIY} \\ 
\textit{IYIIYII} \\ 
\textit{YYYIIYY}
\\ $d=2$  
&
\textit{XXXXXXX} \\ 
\textit{YYIIYYI} \\ 
\textit{YYIYIYI} \\ 
\textit{IYYYYYY} \\ 
\textit{YYIYIIY}
\\ $d=2$  
&
\textit{XXXXXXX} \\ 
\textit{YIIYYYI} \\ 
\textit{YYIIYIY} \\ 
\textit{YYYYIII}
\\ $d=2$  
\tabularnewline
\midrule
$8$ &
\textit{XIXXIIXX} \\ 
\textit{IXIXXXXI} \\ 
\textit{IYYIIYIY} \\ 
\textit{IYYIIIYI} \\ 
\textit{YIYIYIYY} \\ 
\textit{IYYYIYYY} \\ 
\textit{YYYYIIIY}
\\ $d=2$  
&
\textit{XXXXIIXX} \\ 
\textit{XIXXXXXI} \\ 
\textit{IIIIIYYY} \\ 
\textit{IYYYIYYI} \\ 
\textit{YYIIIYII} \\ 
\textit{IYYIYYYY}
\\ $d=2$  
&
\textit{XXIXXXXX} \\ 
\textit{XXXXIXXX} \\ 
\textit{YIIIIYYY} \\ 
\textit{YYIYIIYI} \\ 
\textit{IIYYYIYY}
\\ $d=2$  
\tabularnewline
\midrule
$9$ &
\textit{IIXXIIIXX} \\ 
\textit{XXIXXIIXX} \\ 
\textit{IXXXXIXXI} \\ 
\textit{IIIXXXIIX} \\ 
\textit{IIYYYYIYY} \\ 
\textit{IYYIIYIIY} \\ 
\textit{IYIYYYYIY} \\ 
\textit{YIIIYYYII}
\\ $d=3$  
&
\textit{IXXXXIXIX} \\ 
\textit{XXIIXXXXI} \\ 
\textit{XIXIXXXII} \\ 
\textit{YIIYYIIII} \\ 
\textit{YYIIIYIYY} \\ 
\textit{IIYYIYIYI} \\ 
\textit{YIIYYYYIY}
\\ $d=2$  
&
\textit{XXIIXXXIX} \\ 
\textit{IIXXXXIXX} \\ 
\textit{YYIYYIYII} \\ 
\textit{IYYYIYYIY} \\ 
\textit{IIIYYIIYY} \\ 
\textit{IIIYIYYII}
\\ $d=2$  
\tabularnewline
\midrule
$10$ &
\textit{IXXIXXXIXI} \\ 
\textit{IIXXIIXIXX} \\ 
\textit{IXIXIIXXXI} \\ 
\textit{XIIIXIIXXI} \\ 
\textit{YYIIIIIIYY} \\ 
\textit{YIYYIIIIYY} \\ 
\textit{IYYIIIYYYY} \\ 
\textit{YIYIIYYIYY} \\ 
\textit{IIIYYIIIYI}
\\ $d=3$  
&
\textit{IXIXIIXXXX} \\ 
\textit{IXXXXXXIIX} \\ 
\textit{XIIXXIXXII} \\ 
\textit{YIIYIYYYIY} \\ 
\textit{YIYYIYIIIY} \\ 
\textit{IIYIYIYIIY} \\ 
\textit{YYYIYIIIIY} \\ 
\textit{IYIIIYIIYI}
\\ $d=2$  
&
\textit{XIIXXXIXIX} \\ 
\textit{IIXIXXXIXI} \\ 
\textit{XXIIXIIXXI} \\ 
\textit{YYYIYIIYIY} \\ 
\textit{YIIIIIYIYY} \\ 
\textit{YIYYYYYIII} \\ 
\textit{IIIIYYIYIY}
\\ $d=2$  
\tabularnewline
\midrule
$11$ &
\textit{XIXXIIXIXIX} \\ 
\textit{IXIIXIXXIXX} \\ 
\textit{IIIXXXXIXXI} \\ 
\textit{XXIIXIIXXIX} \\ 
\textit{IIIIIIYYYII} \\ 
\textit{YYIIYIYYYYY} \\ 
\textit{IYYIIIIIIIY} \\ 
\textit{IYYYYIYYIYY} \\ 
\textit{YYYIYIYIYYI} \\ 
\textit{IIYIYYIIIIY}
\\ $d=3$  
&
\textit{IIXXIXXIXXX} \\ 
\textit{XIIXIIXXXII} \\ 
\textit{IIXIXIXIXXI} \\ 
\textit{IXXXIIXXIIX} \\ 
\textit{YIYYIIIYYIY} \\ 
\textit{YYYIIYYIIIY} \\ 
\textit{YYIYYIIIIYI} \\ 
\textit{YIYIIIYIIII} \\ 
\textit{IYYIYIYIYIY}
\\ $d=3$  
&
\textit{IXIIXXXIIIX} \\ 
\textit{XIIXXIXXIXX} \\ 
\textit{XIIIIXXXXIX} \\ 
\textit{XXXIIXXIXXI} \\ 
\textit{YIYYIYYYIII} \\ 
\textit{IIYYIYIYYYY} \\ 
\textit{YYYIYIIYIYI} \\ 
\textit{YIIIYYIYYYI}
\\ $d=3$  
\tabularnewline
\midrule
$12$ &
\textit{XXXIIXIIXXXX} \\ 
\textit{IXIXXXXXIXXI} \\ 
\textit{IIXXIIXXXXII} \\ 
\textit{XXIXXXXIIIII} \\ 
\textit{XIIXIXXXXXXI} \\ 
\textit{IIXXIIIIIXIX} \\ 
\textit{IIYIYIYYYIYY} \\ 
\textit{YYIIYIYIYIYI} \\ 
\textit{YIIIYYYYIIII} \\ 
\textit{IIYYIIYYIIYI} \\ 
\textit{IIYIYYIYYYII}
\\ $d=3$  
&
\textit{XXIXIIXIXXIX} \\ 
\textit{XIXIIIXXXIXI} \\ 
\textit{XXXIIXIXXIXX} \\ 
\textit{IIIXXXIXXIIX} \\ 
\textit{IIYIIIYYIIYY} \\ 
\textit{IYYIYIYIYYYI} \\ 
\textit{IYYYIYIYYYYI} \\ 
\textit{IYIIIIIYIIYY} \\ 
\textit{IIYIIYIIIYYY} \\ 
\textit{YYYYYIIYIIYY}
\\ $d=2$  
&
\textit{XIIXIXXIXXII} \\ 
\textit{IIXXXIXXXIIX} \\ 
\textit{XXXIIIIXXIII} \\ 
\textit{XXXIXIXIIXXI} \\ 
\textit{IYYYYYYYYIII} \\ 
\textit{YIYIIIYYYYII} \\ 
\textit{YIIYIYYYIIIY} \\ 
\textit{YIYYYIIIIIYY} \\ 
\textit{YIYIIIYIIIYI}
\\ $d=3$  
\tabularnewline
\bottomrule
\end{tabular*}
\end{table}

\begin{table}
\caption{\label{tab:AD HC codes CSSY}Generators and distances for the best
CSSY codes found for the AD channel using hill climbing.}

\begin{tabular*}{8.6cm}{@{\extracolsep{\fill}}>{\centering}m{0.5cm}>{\centering}m{2.5cm}>{\centering}m{2.5cm}>{\centering}m{2.5cm}}
\toprule
$n \backslash k$ & $1$ & $2$ & $3$
\tabularnewline
\midrule
$5$ &
\textit{IIXIX} \\ 
\textit{IIIXX} \\ 
\textit{IYYYY} \\ 
\textit{YIYYY}
\\ $d=1$  
&
\textit{XIXXI} \\ 
\textit{XXXII} \\ 
\textit{IYYYY}
\\ $d=1$  
&
\textit{IIYYI} \\ 
\textit{YYYIY}
\\ $d=1$  
\tabularnewline
\midrule
$6$ &
\textit{XXXXXI} \\ 
\textit{XXXXIX} \\ 
\textit{YIIYII} \\ 
\textit{IYIYII} \\ 
\textit{YIYYYY}
\\ $d=2$  
&
\textit{XXXXXX} \\ 
\textit{IXIXXI} \\ 
\textit{IYYYIY} \\ 
\textit{YIIYYY}
\\ $d=2$  
&
\textit{IXXIXI} \\ 
\textit{YIYYYI} \\ 
\textit{YYIIYY}
\\ $d=1$  
\tabularnewline
\midrule
$7$ &
\textit{XXXXIII} \\ 
\textit{XXIIXXI} \\ 
\textit{XIXIIXX} \\ 
\textit{IYIYIYY} \\ 
\textit{YYIIYYI} \\ 
\textit{IIYYYYI}
\\ $d=3$  
&
\textit{IXIIIIX} \\ 
\textit{XXXIXXI} \\ 
\textit{XXXXIIX} \\ 
\textit{IYYYYYY} \\ 
\textit{YYYIYIY}
\\ $d=2$  
&
\textit{XXXXXXX} \\ 
\textit{IIYYYIY} \\ 
\textit{YYIIYIY} \\ 
\textit{IYYIIYY}
\\ $d=2$  
\tabularnewline
\midrule
$8$ &
\textit{IXIXXIII} \\ 
\textit{XXXXXXXI} \\ 
\textit{IXIXIXII} \\ 
\textit{XXIIXXXX} \\ 
\textit{YYIIYYII} \\ 
\textit{YYYIYYYY} \\ 
\textit{IYIYIIIY}
\\ $d=2$  
&
\textit{IXXIIIIX} \\ 
\textit{XXXIXIIX} \\ 
\textit{XXIXIXXX} \\ 
\textit{YYIIYYIY} \\ 
\textit{YYIYYIIY} \\ 
\textit{IIYYIYYY}
\\ $d=2$  
&
\textit{XIXIXXII} \\ 
\textit{IXXIIXXX} \\ 
\textit{XIIXIXXX} \\ 
\textit{YYYYYYIY} \\ 
\textit{IIIIIIYY}
\\ $d=2$  
\tabularnewline
\midrule
$9$ &
\textit{IIXXXIXXX} \\ 
\textit{XXXIIIXII} \\ 
\textit{XIXIXXXIX} \\ 
\textit{IXIXIIXIX} \\ 
\textit{YIIIIYYIY} \\ 
\textit{IYYYIYIII} \\ 
\textit{YIYIYYIII} \\ 
\textit{YIIYYYYYI}
\\ $d=3$  
&
\textit{IXXXXXIIX} \\ 
\textit{XXIXIIIXI} \\ 
\textit{IIXXIIXXX} \\ 
\textit{XIXXIXXXI} \\ 
\textit{YYYIYIIIY} \\ 
\textit{YYIYYYIYI} \\ 
\textit{IYIYIIYII}
\\ $d=2$  
&
\textit{IXXXIXIXX} \\ 
\textit{IIIXXIXXX} \\ 
\textit{XIXIXXIXI} \\ 
\textit{YYYIIYYYI} \\ 
\textit{YYIYIIYYY} \\ 
\textit{IIIYYYIII}
\\ $d=2$  
\tabularnewline
\midrule
$10$ &
\textit{XIIXIXIXXX} \\ 
\textit{IXXXXIIIIX} \\ 
\textit{XIXXXIXIXX} \\ 
\textit{IXXIIIXXXX} \\ 
\textit{YYYIIYIIII} \\ 
\textit{IYIYYYYYIY} \\ 
\textit{IIYIIIYIYY} \\ 
\textit{IIYIYIYYYI} \\ 
\textit{YIYIYYYIII}
\\ $d=3$  
&
\textit{XXXIXXIIXI} \\ 
\textit{XIXXIIXXXI} \\ 
\textit{IXXIIXIXXX} \\ 
\textit{IXXXIIIXII} \\ 
\textit{YYIIYYIYIY} \\ 
\textit{YYYIIIYIYY} \\ 
\textit{YYIYYIYIYI} \\ 
\textit{IYIYYYIIYY}
\\ $d=3$  
&
\textit{XXIXIXXIIX} \\ 
\textit{XXXIXXIIXI} \\ 
\textit{IIIIIXIXXX} \\ 
\textit{XIIXXXIIII} \\ 
\textit{IYIIYYYYYY} \\ 
\textit{YIIYYYYIYI} \\ 
\textit{IIYYIYYIIY}
\\ $d=2$  
\tabularnewline
\midrule
$11$ &
\textit{XIXIIXIXIII} \\ 
\textit{XIIXIXIIIXI} \\ 
\textit{IXXXIIIIXIX} \\ 
\textit{XXXIXXXXIIX} \\ 
\textit{IXIIXXIIXXI} \\ 
\textit{IYIIYIYIIIY} \\ 
\textit{IIIYYIIIIYY} \\ 
\textit{IIYYYYYIIII} \\ 
\textit{YYIYIYIIYYY} \\ 
\textit{YYIIIIIYIYY}
\\ $d=3$  
&
\textit{XIXXXIIXXXI} \\ 
\textit{XXXIIXXIXIX} \\ 
\textit{IIXIIXIXXXX} \\ 
\textit{XIXXIIXXIIX} \\ 
\textit{YIIIIYYIYII} \\ 
\textit{IIYIYYIYIYI} \\ 
\textit{IYIYIIYYYYY} \\ 
\textit{IIIYYYYYYIY} \\ 
\textit{YIYYYYIYYII}
\\ $d=3$  
&
\textit{XXXIIXXXIXI} \\ 
\textit{IXXXXIXXIII} \\ 
\textit{IXIIXXIXXXX} \\ 
\textit{IXXIIXIIIIX} \\ 
\textit{YIYYIYIIIYI} \\ 
\textit{YIYYIIYYIIY} \\ 
\textit{YYYIYYIYIYY} \\ 
\textit{YIIYYIYYYYI}
\\ $d=3$  
\tabularnewline
\midrule
$12$ &
\textit{XIIXIIXXXIII} \\ 
\textit{XXIXIIIIXIIX} \\ 
\textit{XXIIXXXXIXIX} \\ 
\textit{IIXIXIXXXIXX} \\ 
\textit{IIXIXIIIIXIX} \\ 
\textit{IIIXXIXIXXXX} \\ 
\textit{YYYIIIYIIYII} \\ 
\textit{IYIYYYYYYIYY} \\ 
\textit{YYYYIIIYYYYI} \\ 
\textit{YIYYIYYIYIIY} \\ 
\textit{IYYYYYYIIIYI}
\\ $d=3$  
&
\textit{XXXXIIIIIIXI} \\ 
\textit{XXXIXIIIIXIX} \\ 
\textit{XIIXXXXIIXII} \\ 
\textit{XXIXXXIIXXXX} \\ 
\textit{XXXXXIIXXIIX} \\ 
\textit{IIYIYIIYIYYY} \\ 
\textit{YYIIIIIYIYIY} \\ 
\textit{IIIYYYIYYYYI} \\ 
\textit{IYIYYIYYYYIY} \\ 
\textit{YYYYIIYIIYII}
\\ $d=3$  
&
\textit{XXIIIIIXXXXI} \\ 
\textit{XIXIXXIIXIXI} \\ 
\textit{XXIIXXXXXIIX} \\ 
\textit{XXIXIXXIIIXI} \\ 
\textit{YYIYIIIYIIYY} \\ 
\textit{IYIYYIIYYYII} \\ 
\textit{IIYYIIIIIYYI} \\ 
\textit{IYIIYYIIIYIY} \\ 
\textit{YYYIYIYIIYYI}
\\ $d=3$  
\tabularnewline
\bottomrule
\end{tabular*}
\end{table}

\begin{table}
\caption{\label{tab:XZ HC codes lin}Generators and distances for the best
linear codes found for the biased $XZ$ channel using hill climbing.}

\begin{tabular*}{8.6cm}{@{\extracolsep{\fill}}>{\centering}m{0.5cm}>{\centering}m{2.5cm}>{\centering}m{2.5cm}>{\centering}m{2.5cm}}
\toprule
$n \backslash k$ & $1$ & $2$ & $3$
\tabularnewline
\midrule
$5$ &
\textit{XXYIY} \\ 
\textit{ZZXIX} \\ 
\textit{XYIYX} \\ 
\textit{ZXIXZ}
\\ $d=3$  
&
-
&
-
\tabularnewline
\midrule
$6$ &
-
&
\textit{YIXIYX} \\ 
\textit{XIZIXZ} \\ 
\textit{XYZZZY} \\ 
\textit{ZXYYYX}
\\ $d=2$  
&
-
\tabularnewline
\midrule
$7$ &
\textit{ZIZXZZY} \\ 
\textit{YIYZYYX} \\ 
\textit{IYYYZII} \\ 
\textit{IXXXYII} \\ 
\textit{ZYYIIXI} \\ 
\textit{YXXIIZI}
\\ $d=3$  
&
-
&
\textit{IXZYXXZ} \\ 
\textit{IZYXZZY} \\ 
\textit{YYXZXIY} \\ 
\textit{XXZYZIX}
\\ $d=2$  
\tabularnewline
\midrule
$8$ &
-
&
\textit{IYXZYYXI} \\ 
\textit{IXZYXXZI} \\ 
\textit{XIYXYYIY} \\ 
\textit{ZIXZXXIX} \\ 
\textit{IXYYYXIY} \\ 
\textit{IZXXXZIX}
\\ $d=3$  
&
-
\tabularnewline
\midrule
$9$ &
\textit{ZZIIIXIIY} \\ 
\textit{YYIIIZIIX} \\ 
\textit{IXZZZXZZX} \\ 
\textit{IZYYYZYYZ} \\ 
\textit{IYIZZIZXX} \\ 
\textit{IXIYYIYZZ} \\ 
\textit{IIYYIIXXI} \\ 
\textit{IIXXIIZZI}
\\ $d=3$  
&
-
&
\textit{ZXXXYXXIZ} \\ 
\textit{YZZZXZZIY} \\ 
\textit{ZZZXIYXZY} \\ 
\textit{YYYZIXZYX} \\ 
\textit{ZYZIIZZXI} \\ 
\textit{YXYIIYYZI}
\\ $d=3$  
\tabularnewline
\midrule
$10$ &
-
&
\textit{XZYXXXYZZX} \\ 
\textit{ZYXZZZXYYZ} \\ 
\textit{ZYXYXZZXXX} \\ 
\textit{YXZXZYYZZZ} \\ 
\textit{YXIZYYXIYZ} \\ 
\textit{XZIYXXZIXY} \\ 
\textit{ZIYYYXYZYI} \\ 
\textit{YIXXXZXYXI}
\\ $d=3$  
&
-
\tabularnewline
\midrule
$11$ &
\textit{YZXIIYYZIZX} \\ 
\textit{XYZIIXXYIYZ} \\ 
\textit{YYXZYXIZYYY} \\ 
\textit{XXZYXZIYXXX} \\ 
\textit{YIXYYXZXZZY} \\ 
\textit{XIZXXZYZYYX} \\ 
\textit{IZZYXIYIYXX} \\ 
\textit{IYYXZIXIXZZ} \\ 
\textit{ZXZXIZIIZXZ} \\ 
\textit{YZYZIYIIYZY}
\\ $d=3$  
&
-
&
\textit{XIZZZIYYIXY} \\ 
\textit{ZIYYYIXXIZX} \\ 
\textit{XYXIYIYXYIZ} \\ 
\textit{ZXZIXIXZXIY} \\ 
\textit{XXZXXIIZYYI} \\ 
\textit{ZZYZZIIYXXI} \\ 
\textit{ZIYIYYYYIYY} \\ 
\textit{YIXIXXXXIXX}
\\ $d=3$  
\tabularnewline
\midrule
$12$ &
-
&
\textit{IIIIYYIIIIZZ} \\ 
\textit{IIIIXXIIIIYY} \\ 
\textit{YYZYZYIIYIIY} \\ 
\textit{XXYXYXIIXIIX} \\ 
\textit{IZZZZZZIXXYY} \\ 
\textit{IYYYYYYIZZXX} \\ 
\textit{ZXYIYYXYIZII} \\ 
\textit{YZXIXXZXIYII} \\ 
\textit{ZYYXZZIIZXZZ} \\ 
\textit{YXXZYYIIYZYY}
\\ $d=4$  
&
-
\tabularnewline
\bottomrule
\end{tabular*}
\end{table}

\begin{table}
\caption{\label{tab:AD HC codes lin}Generators and distances for the best
linear codes found for the AD channel using hill climbing.}

\begin{tabular*}{8.6cm}{@{\extracolsep{\fill}}>{\centering}m{0.5cm}>{\centering}m{2.5cm}>{\centering}m{2.5cm}>{\centering}m{2.5cm}}
\toprule
$n \backslash k$ & $1$ & $2$ & $3$
\tabularnewline
\midrule
$5$ &
\textit{XIXZZ} \\ 
\textit{ZIZYY} \\ 
\textit{IXZXZ} \\ 
\textit{IZYZY}
\\ $d=3$  
&
-
&
-
\tabularnewline
\midrule
$6$ &
-
&
\textit{IXIXYY} \\ 
\textit{IZIZXX} \\ 
\textit{YXXIIY} \\ 
\textit{XZZIIX}
\\ $d=2$  
&
-
\tabularnewline
\midrule
$7$ &
\textit{IXYYXYZ} \\ 
\textit{IZXXZXY} \\ 
\textit{XXIYYYY} \\ 
\textit{ZZIXXXX} \\ 
\textit{XZXYIZY} \\ 
\textit{ZYZXIYX}
\\ $d=3$  
&
-
&
\textit{XZZZIXY} \\ 
\textit{ZYYYIZX} \\ 
\textit{IXYZYXY} \\ 
\textit{IZXYXZX}
\\ $d=2$  
\tabularnewline
\midrule
$8$ &
-
&
\textit{IYIYZYXX} \\ 
\textit{IXIXYXZZ} \\ 
\textit{ZIIXYXYX} \\ 
\textit{YIIZXZXZ} \\ 
\textit{ZZZXXIIZ} \\ 
\textit{YYYZZIIY}
\\ $d=3$  
&
-
\tabularnewline
\midrule
$9$ &
\textit{YXXIXZXZY} \\ 
\textit{XZZIZYZYX} \\ 
\textit{ZXYYXYIZZ} \\ 
\textit{YZXXZXIYY} \\ 
\textit{IZYYIYXIX} \\ 
\textit{IYXXIXZIZ} \\ 
\textit{IIIIYIZYZ} \\ 
\textit{IIIIXIYXY}
\\ $d=3$  
&
-
&
\textit{ZZIXIZIXY} \\ 
\textit{YYIZIYIZX} \\ 
\textit{YZIXXXXYX} \\ 
\textit{XYIZZZZXZ} \\ 
\textit{IXYXYXXZZ} \\ 
\textit{IZXZXZZYY}
\\ $d=3$  
\tabularnewline
\midrule
$10$ &
-
&
\textit{XYYIIYZZYX} \\ 
\textit{ZXXIIXYYXZ} \\ 
\textit{ZXYIYIXXZZ} \\ 
\textit{YZXIXIZZYY} \\ 
\textit{XXIIYZZXYY} \\ 
\textit{ZZIIXYYZXX} \\ 
\textit{YXXYYXXIIZ} \\ 
\textit{XZZXXZZIIY}
\\ $d=3$  
&
-
\tabularnewline
\midrule
$11$ &
\textit{YIYZYZZYXXY} \\ 
\textit{XIXYXYYXZZX} \\ 
\textit{YXIIIYZZIXI} \\ 
\textit{XZIIIXYYIZI} \\ 
\textit{XYXZIZYIXIZ} \\ 
\textit{ZXZYIYXIZIY} \\ 
\textit{XIZXIXXXZIX} \\ 
\textit{ZIYZIZZZYIZ} \\ 
\textit{YXXZXYYIXYX} \\ 
\textit{XZZYZXXIZXZ}
\\ $d=3$  
&
-
&
\textit{XZXYYIIXIZZ} \\ 
\textit{ZYZXXIIZIYY} \\ 
\textit{YIIXIZYXXII} \\ 
\textit{XIIZIYXZZII} \\ 
\textit{ZIIIXZZYZZY} \\ 
\textit{YIIIZYYXYYX} \\ 
\textit{IYXZZZIZXIZ} \\ 
\textit{IXZYYYIYZIY}
\\ $d=3$  
\tabularnewline
\midrule
$12$ &
-
&
\textit{IIIIYXYYXZXX} \\ 
\textit{IIIIXZXXZYZZ} \\ 
\textit{IIZIXZZZIZZY} \\ 
\textit{IIYIZYYYIYYX} \\ 
\textit{IZIZIXYIXZII} \\ 
\textit{IYIYIZXIZYII} \\ 
\textit{ZIYXIXXIIIYI} \\ 
\textit{YIXZIZZIIIXI} \\ 
\textit{XIYIYIXXIYZX} \\ 
\textit{ZIXIXIZZIXYZ}
\\ $d=4$  
&
-
\tabularnewline
\bottomrule
\end{tabular*}
\end{table}

\endgroup

\FloatBarrier
\end{document}